\documentclass{emulateapj}
\usepackage{epstopdf}
\usepackage{color}
\newcommand{\beq}{\begin{equation}}
\newcommand{\eeq}{\end{equation}}
\newcommand{\bea}{\begin{eqnarray}}
\newcommand{\eea}{\end{eqnarray}}

\begin{document}
\title{The Kinetic Sunyaev-Zel'dovich effect as a probe of
the physics of cosmic reionization: the effect of
self-regulated reionization}

\author{Hyunbae Park\altaffilmark{1}}
\author{Paul R. Shapiro\altaffilmark{1}}
\author{Eiichiro Komatsu\altaffilmark{1,2,3}}
\author{Ilian T. Iliev\altaffilmark{4}}
\author{Kyungjin Ahn\altaffilmark{5}}
\author{Garrelt Mellema\altaffilmark{6}}

\altaffiltext{1}{Texas Cosmology Center and the Department of Astronomy, The University of Texas at Austin, 1 University Station, C1400, Austin, TX 78712, USA}
\altaffiltext{2}{Kavli Institute for the Physics and Mathematics of the Universe, Todai Institutes for Advanced Study, the University of Tokyo, Kashiwa, Japan 277-8583 (Kavli IPMU, WPI)}
\altaffiltext{3}{Max-Planck-Institut f\"{u}r Astrophysik, Karl-Schwarzschild Str. 1, 85741 Garching, Germany}
\altaffiltext{4}{Astronomy Centre, Department of Physics and Astronomy, Pevensey II Building, University of Sussex, Falmer, Brighton BN1 9QH}
\altaffiltext{5}{Department of Earth Science Education, Chosun University, Gwangju 501-759, Korea}
\altaffiltext{6}{Stockholm Observatory, AlbaNova University Center, Stockholm University, SE-106 91 Stockholm, Sweden}

\begin{abstract}
We calculate the angular power spectrum of the Cosmic Microwave Background (CMB) temperature fluctuations induced by the kinetic Sunyaev-Zel'dovich (kSZ) effect from the epoch of reionization (EOR). We use detailed $N$-body+radiative transfer simulations to follow
 inhomogeneous reionization of the intergalactic medium (IGM).  
For the first time we take into account the 
``self-regulation'' of reionization: star
 formation in low-mass dwarf galaxies $(10^8~M_\sun \lesssim M \lesssim 10^9~M_\sun)$ or
 minihalos $(10^5~M_\sun \lesssim M \lesssim 10^8~M_\sun)$  is
 suppressed if these halos form in
the regions that were already ionized or Lyman-Werner
 dissociated. 
 Some previous work suggested that the amplitude of the kSZ power
 spectrum from the EOR can be 
 described by a two-parameter family: the epoch of half ionization and
 the duration of reionization. However, we argue that this picture
 applies only to simple forms of the reionization history which are
 roughly symmetric about the half-ionization epoch.
 In self-regulated reionization, the universe begins to be ionized early,
 maintains a low level of ionization for an extended period, and then finishes
 reionization as soon as high-mass atomically-cooling halos
 dominate. While inclusion of self-regulation affects the amplitude
 of the kSZ power spectrum only modestly ($\sim 10\%$), it can change
 the duration of reionization by a factor of more than two. 
 We conclude that the simple two-parameter family does not capture the
 effect of a  physical, yet complex, reionization history caused by
 self-regulation.
 When added to the post-reionization kSZ contribution,
our prediction for the total kSZ power spectrum is below the current upper bound  
 from the South Pole Telescope.
Therefore, the current upper bound on the kSZ effect from the EOR is consistent
 with our understanding of the physics of reionization.
\end{abstract}

\section{Introduction}

How was the intergalactic medium (IGM) reionized before $z=6$?
The secondary anisotropy of the cosmic microwave background (CMB) 
at $l>3000$ allows us to probe the physics of cosmic
reionization via the kinetic Sunyaev-Zel'dovich effect
\citep[kSZ;][]{sunyaev/zeldovich:1980}. The temperature of the CMB changes
as free electrons in ionized gas Compton scatter CMB photons: the bulk peculiar velocity
of electrons induces Doppler shifts in the energy of the CMB photons. While the
spectrum of the CMB remains that of a black body, its temperature
changes.\footnote{A related effect results from the thermal motions of free electrons in the hot intracluster gas, called the thermal SZ effect (tSZ). Multiwavelength observations allow a distinction between the kSZ and tSZ effects on the CMB. Here, we shall focus on the kSZ signal alone.}

Inhomogeneity in the density and velocity of electrons, as well as
inhomogeneity in ionization fraction, will induce temperature 
fluctuations in the CMB, $\Delta T/T$, given by
\beq \label{Eq:kSZ}
\frac{\Delta T(\hat{\gamma})}{T} = -\int d\tau e^{-\tau} \frac{\hat{\gamma}\cdot\bold{v}}{c},
\eeq
where $\hat{\gamma}$ is the line-of-sight unit vector, $\bold{v}$ the
peculiar velocity field, and $\tau$ the optical depth to Thomson scattering integrated through the IGM from $z=0$ to the surface of last scattering at $z_{\rm{rec}}\approx 10^3$, where 
\beq
d\tau = c~n_e (z) \sigma_{\rm{T}}(\frac{dt}{dz})dz .
\eeq
There are two contributions to the kSZ signal:
\begin{itemize}
\item[1.] {\bf Post-reionization contribution}. This is the contribution
	  from redshifts below $z = z_{\rm{ov}}$, where $z_{\rm ov}$ is 
	  the redshift at which reionization is finished, 
	  when individual H II bubbles fully overlap with one another.
	  While the post-reionization contribution depends
	  upon the value of $z_{\rm{ov}}$, for which quasar absorption
	  spectra suggest $z_{\rm{ov}} \sim 6-7$, it is not too
	  sensitive to the exact value of $z_{\rm{ov}}$. We shall not
	  discuss this contribution in this paper, but discuss it in a
	  subsequent paper (Park et al., in preparation).
\item[2.] {\bf Reionization contribution}. This is the
	  contribution from redshifts above $z=z_{\rm ov}$, where the
	  ionization was patchy and incomplete. This contribution
	  depends not only on $z_{\rm{ov}}$, but also on the details of
	  the time and spatial variation of inhomogeneous reionization,
	  which are not yet well constrained; thus, we must
	  explore how predictions vary for
	  different models of reionization. The reionization
	  contribution is the main
	  focus of this paper.
\end{itemize}

Modeling the reionization contribution is a challenge, as the universe
was not ionized homogeneously, but in patches.
These patches grow over time until they overlap,
finishing reionization of the universe. The distribution of
these patches is determined by non-linear physics: non-linear clustering
of the sources of ionizing photons; non-linear clumping of gas in the
IGM; and complex 
morphologies of patches resulting from propagation of ionization fronts in
the clumpy IGM. Accurately calculating the reionization contribution
thus requires numerical simulations of cosmological structure
formation coupled with radiative transfer.

To model the formation and spatial clustering of the sources of ionizing
photons, cosmological simulations must be performed in a volume large
enough to capture the crucial spatial variations of this process in a statistically meaningful way.
This requires a volume greater than $\sim 100~ $comoving Mpc across, because H II bubbles can typically grow as large as $\sim 20~$comoving Mpc in size. These
simulations must also have a high enough mass resolution to resolve the
formation of the individual galaxies which are the sources of ionizing radiation; 
thus, billions of particles are required. 
The radiative transfer of ionizing photons is then
calculated on the IGM density and velocity fields computed by the
cosmological simulation.

What do current observational data tell us? 
The South Pole Telescope (SPT) experiment 
has detected an excess temperature anisotropy
for the CMB on small angular scales, which they attribute to the SZ effect.
By subtracting the dominant contribution from the tSZ effect by using multiwavelength
observations to distinguish it from the kSZ effect, the SPT detection yields an upper limit to
the total kSZ contribution.\footnote{The post-reionization kSZ effect due to the pairwise relative motions of galaxy clusters has been detected by the Atacama Cosmology Telescope \citep{2012PhRvL.109d1101H}, but its contribution to the CMB temperature fluctuation power spectrum has not been detected yet.}
The measurements are usually reported in terms of the angular power
spectrum, $C_l\equiv \frac1{2l+1}\sum_m|a_{lm}|^2$. Here, $a_{lm}\equiv \int
d^2\hat{\gamma}\Delta T(\hat{\gamma})Y_{lm}^*(\hat{\gamma})$ 
is the coefficient of spherical-harmonics mode, 
$Y_{lm}$, of $\Delta T$.
The SPT collaboration reports
their measurements in terms of the quantity 
\begin{equation}
D_l\equiv \frac{l(l+1)C_l}{2\pi},
\end{equation}
which we shall compute in this paper. SPT has placed
an upper bound on the kSZ $D_l$ at $l=3000$ of $D^{\rm{kSZ}}_{l=3000}<
2.8 ~\mu K^2$ \citep{reichardt/etal:2011}. 
The detection of the total SZ effect is complicated by the possible contamination of the fluctuating signal caused by the cosmic infrared background (CIB) from individual galaxies.
The kSZ limit loosens to $6.0~\mu K^2$ 
when allowance is made for a possible correlation between the thermal
Sunyaev-Zel'dovich effect \citep[tSZ;][]{zeldovich/sunyaev:1969} and the CIB. 
Our goal is to see whether these current upper bounds are consistent with our models of reionization.

Following the early analytical calculation done by Vishniac for
linear density 
and velocity perturbations in a fully ionized medium
\citep{vishniac:1987,jaffe/kamionkowski:1998},
calculations of the kSZ effect by cosmic reionization have steadily improved over time.
Further analytical calculations later incorporated the effects of inhomogeneous
reionization in an approximate manner
\citep{gruzinov/hu:1998,santos/etal:2003}.
A ``semi-numerical'' approach
was also developed by combining the simulated density and velocity
fields from N-body simulations
with an analytical ansatz for tracking the reionization process
\citep{zahn/etal:2005,mcquinn/etal:2005}. 

Early, pioneering
calculations using structure formation simulations coupled with radiative transfer 
to model inhomogeneous reionization numerically \citep{gnedin/jaffe:2001,salvaterra/etal:2005} 
underestimated the amplitude of the kSZ signal, as they used computational boxes 
too small to capture the impact of large-scale velocity modes and H II bubbles or an accurate measure of the duration of the global EOR.
This was demonstrated by
the first calculations of reionization based on truly large-scale ($>100$~Mpc) radiative transfer
simulations, which resolved the formation of all galactic halo sources above 
$2\times 10^9 M_\sun$ \citep{iliev/etal:2007,iliev/etal:2008}.
These later simulations demonstrated the importance of a
large enough simulation volume to capture the effects of long-wavelength
fluctuations properly. 
They were also the first to realize
that it is necessary to correct the kSZ power spectrum for the missing velocity
power due to the finite box size of the simulations. 

For the mass range of galactic halos resolved by these simulations, $\gtrsim 10^9 M_\sun$, stars -- the {\it sources} of reionization -- were able to form when the primordial composition gas inside the halos cooled radiatively by atomic processes involving H atoms.
They are known as ``atomic cooling halos'' to distinguish them from minihalos of mass $M\lesssim 10^8 M_\sun$, with virial temperature $T_{\rm{vir}}\lesssim10^4 K$, for which star formation is possible only if $\rm{H}_2$ molecules form in sufficient abundance to cool the gas below $T_{\rm{vir}}$ by rotational-vibrational line excitation. Atomic-cooling halos with $10^8 M_\sun \lesssim M \lesssim 10^9 M_\sun$ also exist and are even more abundant than those with $M\gtrsim 10^9 M_\sun$. These low-mass atomic-cooling halos (``LMACHs''), however, are prevented from forming stars if they form within an ionized patch of the IGM, where the gas pressure of the photoheated IGM opposes the accretion of baryons onto these halos. This  ``self-regulates'' their contribution to reionization as the global ionized fraction grows with time and more and more of these halos are born within the ionized zones \citep{1994ApJ...427...25S,iliev/etal:2007b}. While the precise value of halo mass which defines the upper edge of this ``Jeans-filtered''  mass-range is still uncertain, the high-mass atomic-cooling halos (``HMACHs'') above $\sim 10^9 M_\sun$ are generally free of this suppression.

To simulate the impact of both LMACHs and HMACHs on reionization, 
it was necessary for \citet{iliev/etal:2007b} to increase their halo mass resolution so as to resolve all the LMACHs, too, by reducing the simulation box size to 53 Mpc on a side at fixed N-body particle number. This led to the first radiative transfer simulations of ``self-regulated'' reionization, which demonstrated the importance of including and then suppressing the LMACHs to start reionization earlier and extend its duration \citep{iliev/etal:2007b}. While the end of reionization is still set by the rapid rise of the HMACHs, in that case, when they eventually surpass the saturated contribution of the suppressible LMACHs, the effect of the LMACHs is to boost the electron-scattering optical depth, $\tau$, integrated through the EOR. Such an effect can be important for the kSZ fluctuations from the EOR, too, but simulating this required us to increase the simulation volume again while retaining the high mass resolution required to resolve the LMACHs, too.

Our next generation of simulations involved boxes 163 Mpc on a side, 
a volume large enough to predict observables like the kSZ effect,
but with N-body simulations large enough to resolve
all halos down to $10^8 M_\sun$ and incorporate ionization
suppression (``Jeans-filtering'') of the halos of mass between
$10^8M_\sun$ and $10^9M_\sun$ \citep{iliev/etal:2011}. 
These smaller-mass halos (LMACHs) are more
abundant and likely to be more efficient ionizing sources, as 
they may have higher escape fraction and emissivity \citep{iliev/etal:2011}.
However, as described above, they may
be suppressed as sources if they form inside ionized regions, where
ionization heats the gas and makes its pressure high enough to resist
gravitational collapse into such small galaxies. 
Recently, an additional simulation was performed, including this new physics,
in an even larger volume $(\sim 600~\rm{Mpc})$ (Iliev et al. in preparation).

\cite{ahn/etal:2012} expanded the mass range even further by accounting
for starlight emitted by minihalos ($10^5-10^8~M_\sun$), as well.
 In addition to their Jeans-mass filtering in ionized regions, 
they may also be suppressed if molecular hydrogen
in minihalos is photo-dissociated by Lyman-Werner band photons 
in the UV background below 13.6 eV also emitted by the sources of reionization. We thus have a simulated model which takes into account all the halos 
down to $10^5~M_\sun$ as sources of reionization.

It is important now to determine if and how the kSZ fluctuations from the epoch of
reionization are different from the previous predictions when this
``self-regulated'' reionization is taken into account. That is the prime
focus of this paper. Some of our results were first summarized in \citet{2012AIPC.1480..248S}.

Recently,
\cite{mesinger/mcquinn/spergel:2011}, \cite{zahn/etal:2012} and \cite{battaglia/etal:2012b} 
compared the predicted kSZ power spectra from their semi-numerical
calculations, to the upper bounds from the SPT data
\citep{reichardt/etal:2011}, obtaining limits on
the epoch and the duration of the reionization. 
Those studies concluded that, for a given
value of the total Thomson-scattering optical depth, the reionization
contribution to the kSZ signal is mostly sensitive to the duration of
the reionization defined as $\Delta z \equiv z_{99\%} - z_{20\%}$
\citep{zahn/etal:2012} or
$z_{75\%}-z_{25\%}$ \citep{mesinger/mcquinn/spergel:2011,battaglia/etal:2012b}. 
\cite{zahn/etal:2012} claim that the upper bound on $D^{\rm
kSZ}_{l=3000}$ from the SPT data implies $\Delta z < 4$ (95\%~CL) for
no tSZ-CIB correlation, and $\Delta z < 7$ (95\%~CL) for the maximum possible
tSZ-CIB correlation. However, as their methods are based on an
analytical ansatz for the reionization process, 
it is necessary to use more self-consistent calculations of radiative transfer such as our simulation results to revisit this issue.
We note that \cite{zahn/etal:2011} compared their semi-numerical
approach to their own numerical simulations using radiative transfer,
finding an agreement at the level of 50\%.

The remainder of this paper is organized as follows. 
In Section 2, we express the kSZ power spectrum in terms of a
line-of-sight integral of the transverse momentum power spectrum, and
show how the transverse momentum power spectrum is related to the
statistics of the density and velocity fields of ionized gas.
In Section 3, we describe the details of the simulations used for our
study. In Section 4, we present our predictions for the kSZ
power spectrum and discuss the effects of inhomogeneous reionization as
well as of self-regulated reionization.
In Section 5, we compare our results with the recent semi-numerical
calculations, and show that inclusion of self-regulated
reionization qualitatively changes the parameter dependence of the kSZ 
power spectrum from that without self-regulation. In Section 6, we summarize our conclusions. 
In Appendix~A, we give the derivation of the kSZ power
spectrum written in terms of the transverse momentum power spectrum. In
Appendix~B, we show how to correct for the missing power due to a finite
box size of simulations in our method.

\section{Basics}

\begin{figure*}
  \begin{center}
    \includegraphics[scale=0.6]{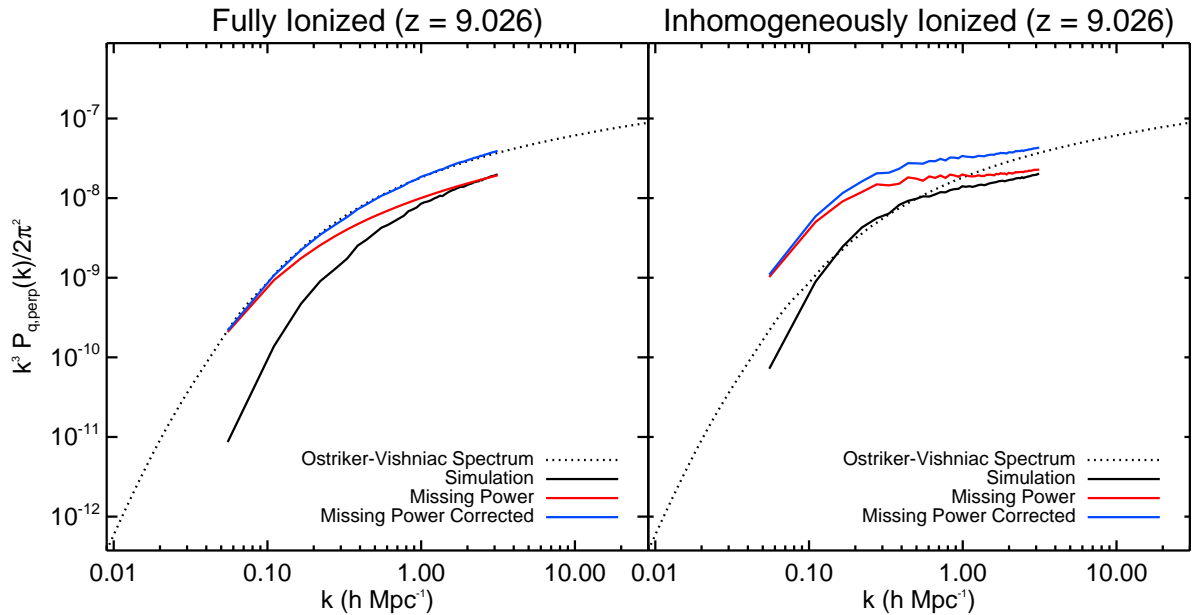}
  \caption{Dimensionless power spectra of the curl of the momentum
   field, $k^3P_{q_\perp}(k)/(2\pi^2)$, at $z=9$ calculated from the
   simulation with $114~h^{-1}~{\rm Mpc}$ in a side. The black solid lines
   show the raw power spectrum obtained from the $N$-body simulation, while the
   blue lines show the power spectrum after being corrected for the
   missing velocity power due to a finite box size of the
   simulation. The red lines show the missing power added to the black
   solid lines. The dotted lines show the analytical OV spectrum given in
   Equation~(\ref{OV}). Left: fully ionized case. An excellent agreement
   between the OV spectrum  and the corrected power spectrum shows the
   validity of our simulation as well as that of our method to correct
   for the missing velocity  power. Right: inhomogeneously ionized
   case, L3. The power spectrum is significantly enhanced at $k\lesssim
   1~h~{\rm Mpc}^{-1}$.}
  \label{fig:missing power corrected Pq}
  \end{center}
\end{figure*}

\subsection{Angular power spectrum of the kSZ effect}

As the Thomson-scattering optical depth, $\tau$, is proportional to the
free electron number density, the kSZ effect given by
Equation~(\ref{Eq:kSZ}) depends mainly on the {\it specific ionized momentum} 
field of the ionized medium, 
\begin{equation}
\bold{q} \equiv \chi \bold{v}(1+\delta),
\end{equation}
henceforth referred to only as ``momentum''.
Here, $\chi
\equiv n_e/(n_{\rm{H}}+2n_{\rm{He}})$ is the ionization
fraction, 
and $\delta\equiv (\rho - \bar\rho)/{\bar\rho}$ is the density contrast
of baryons. In general, the baryon density is
different from the dark matter density, especially on
scales smaller than the Jeans length. In this paper, we shall assume
that baryons trace dark matter particles, as we are interested in scales
bigger than the Jeans length of gas at $10^4$~K.

We rewrite Equation~(\ref{Eq:kSZ}) using $\bold{q}$ as
\beq \label{Eq:kSZ2}
\frac{\Delta T}{T}(\hat{\gamma}) = -\frac{\sigma_T \bar{n}_{e,0}}{c}\int \frac{ds}{a^2} ~ e^{-\tau} \bold{q}\cdot \hat{\gamma}.
\eeq
Here, $\sigma_T$ is the Thomson scattering cross section, $\bar{n}_{e,0}
= \bar{n}_{\rm{H,0}}+2\bar{n}_{\rm{He,0}}$, is the mean number density
of electrons at the (fully-ionized) present epoch, and $s$ is the distance travelled by
photons from a source to the observer in comoving units. 

The kSZ angular power spectrum is given by\footnote{
All previous numerical calculations of the kSZ power
spectrum first created maps using Equation~(\ref{Eq:kSZ2}) and then
measured $C_l$ from the two-dimensional Fourier transform of the
simulated maps. In this paper, we shall use Equation~(\ref{C_l}) to
compute $C_l$ using $P_{q_\perp}$ measured from three-dimensional
simulation boxes at various redshifts, without ever creating
maps. While we are the first to apply this method to the
computation of 
the kSZ power spectrum, this method has been applied successfully to the
computation of the tSZ power spectrum \citep{refregier/etal:2000} as
well as to that of the power spectrum of anisotropy of the near infrared
background
\citep{fernandez/etal:2010,fernandez/etal:2012}. 
} 
\citep[See Appendix~\ref{app:cl} for derivation; also see][but note that their Equation~(4)
contains a typo: it is off by a factor of $(c/H_0)^2$]{ma/fry:2002}:
\beq \label{C_l}
C_l = 
\left(\frac{\sigma_T \bar{n}_{e,0}}{c}\right)^2
\int\frac{ds}{s^2a^4} e^{-2\tau}
\frac{P_{q_{\perp}}(k=l/s,s)}{2},
\eeq
where $\tilde\bold{q}_{\perp}(\bold{k}) = \tilde\bold{q}(\bold{k}) -
\hat{k}[\tilde\bold{q}(\bold{k})\cdot\hat{k}]$ is the
projection of 
$\tilde\bold{q}(\bold{k})\equiv \int d^3\bold{x}~e^{i\bold{k}\cdot
\bold{x}} \bold{q}(\bold{x})$ on the plane perpendicular to the the mode
vector $\bold{k}$ (i.e., $\tilde{\bold{q}}_\perp\cdot\bold{k}=0$), $\hat{k}\equiv \bold{k}/|\bold{k}|$ is a unit
vector, and $P_{q_\perp}$ is the power spectrum of
$\tilde\bold{q}_\perp$ defined by $(2\pi)^3 P_{q_\perp}(k)\delta^D
(\bold{k} - \bold{k}^\prime) \equiv
\langle\tilde\bold{q}_\perp(\bold{k})\cdot\tilde\bold{q}_\perp^*
(\bold{k}^\prime)\rangle$. Note that $\tilde\bold{q}_{\perp}$ is often
called a transverse (or curl) mode. A longitudinal (or gradient)
mode is parallel to $\bold{k}$  and is given by
$\tilde\bold{q}_{\parallel}(\bold{k}) =
\hat{k}[\tilde\bold{q}(\bold{k})\cdot\hat{k}]$. 

As we show in Appendix~\ref{app:cl}, in the small-angle
approximation, 
the line-of-sight integral cancels out the contribution from
$\tilde\bold{q}_{\parallel}$ and a half of the power of
$\tilde\bold{q}_{\perp}$, leaving only the remaining half of
$P_{q_{\perp}}$. This explains a factor of two in the
denominator of Equation~(\ref{C_l}).

Helium atoms are assumed to be
singly ionized where hydrogen atoms are ionized at least until $z_{\rm ov}$, 
the redshift which all the H II bubbles overlap to finish the ionization of hydrogen atoms due to the similar ionization potential of H I and He I. 
Helium atoms remain singly ionized until much later,
$z\approx 3$, after which they are thought to be doubly ionized. 
As we are interested only in the epoch of hydrogen-reionization,
$z\gtrsim 6$,
we shall assume that the ionized fraction, $\chi$, is given by
$\chi=(0.92)X$, where $X$ is the hydrogen ionized fraction at each point
in our radiative transfer simulation: $\chi$ saturates at 0.92 in fully
ionized regions during hydrogen reionization, as 8\% of the electrons
are left bound in singly-ionized helium atoms.

\subsection{Power spectrum of the curl of the momentum} \label{Sec. momentum power}

\begin{table*}
\caption{Reionization simulation parameters and global reionization history results}
\begin{center}
    \begin{tabular}{|l|l|l|l|l|l|l|l|}
\hline

        Label    &$g_{\gamma,\rm{H}}~(f_{\gamma,\rm{H}})$&$g_{\gamma,\rm{L}}~(f_{\gamma,\rm{L}})$&$g_{\gamma,\rm{MH}}~(f_{\gamma,\rm{MH}})\footnote{MH efficiencies $g_{\gamma,\rm{MH}}~(f_{\gamma,\rm{MH}})$
                    quoted here are for the minimum-mass halo assumed to contribute, $10^5~M_\sun$,
                    which is roughly comparable to the average value for the minihalos 
                    integrated over the halo mass function. 
                    The efficiency of any MH of a given mass $M$ is obtained simply by multiplying 
                    to the quoted 
                    $g_{\gamma,\rm{MH}}~(f_{\gamma,\rm{MH}})$ by $(\frac{10^5~M_\sun}{M})$.
                    }          
$&$\tau_{\rm{es}}$ & $z_{10\%}$  & $z_{90\%}$ & $z_{\rm{ov}}$  \\ \hline
        L1        & 8.7(10)  & 130(150) & -          & 0.080 & 13.3 & 8.6 & 8.3      \\  
        L2(XL2)        & 1.7(2)  & 8.7(10)  & -          & 0.058 & 9.9   & 6.9 & 6.7      \\ 
        L2M1J1   & 1.7(2)  & 8.7(10)  & 5063(1030) & 0.086 & 17.4   & 6.9 & 6.7      \\  
        L3        & 21.7(25)& -     & -         & 0.070 & 10.3 & 9.1 & 8.4      \\  \hline
    \end{tabular}
\end{center}
\label{RT parameters}
\end{table*}

Our goal is to compute the power spectrum of the curl of the momentum
field, $P_{q_{\perp}}$, and evaluate Equation~(\ref{C_l}) to obtain $C_l$.

Assuming that the velocity field stays longitudinal, i.e.,
parallel to $\bold{k}$, $P_{q_{\perp}}$ is given by
the second-order term in the momentum: $\bold{q}_{\perp} =
(\int\frac{d^3k^{\prime}}{(2\pi)^3}\delta(\bold{k}-\bold{k}^{\prime})\bold{v}(\bold{k}^{\prime}))_{\perp}$. This assumption is exact in the linear regime and is approximately true in 
the non-linear regime, as this second-order term dominates in the
non-linear regime anyway. This gives \citep{ma/fry:2002}
\begin{eqnarray} 
\nonumber
P_{q_{\perp}} (k,z) &=& \int \frac{d^3k^\prime}{(2\pi)^3}
 (1-{\mu^\prime}^2)\left[
P_{\delta\delta}(|\bold{k}-\bold{k^\prime}|)P_{vv}(k^\prime)\right.\\
\label{MF}
& &\qquad\left.-\frac{k^\prime}{|\bold{k}-\bold{k^\prime}|}P_{\delta v}(|\bold{k}-\bold{k^\prime}|)P_{\delta v}(k^\prime)\right],
\end{eqnarray}
where $\mu^{\prime}\equiv \hat{k}\cdot\hat{k}^{\prime}$. Here,
$P_{\delta\delta}P_{vv}$ term gives a positive contribution, 
whereas $P_{\delta
v}P_{\delta v}$ term gives a negative contribution from the density
field correlated with the velocity field that does not have a curl
component. 

Due to a finite
box size of simulations, we must correct for the missing velocity
power coming from modes whose wavelength is longer than the size of the
simulation box \citep{iliev/etal:2007}. We shall describe our correction method in Appendix~\ref{Sec. Correcting for the Missing Power in Simulation}.

At high redshift where the density and velocity
fields are still in the linear regime, the velocity power spectrum is
related to the linear density power spectrum by $P_{vv}(k) =
(\dot{a}f/k)^2 P^{\rm{lin}}_{\delta\delta}(k)$, where $f\equiv
d\ln\delta/d\ln a$ and $a(t)$ is the Robertson-Walker scale factor. This
gives the so-called 
Ostriker-Vishniac (OV) spectrum \citep{vishniac:1987}:
\begin{eqnarray}
\nonumber
P^{\rm OV}_{q_{\perp}}(k,z) &=& \dot{a}^2f^2\int
 \frac{d^3k^\prime}{(2\pi)^3} P_{\delta\delta}^{\rm{lin}} (|\bold{k} -
 \bold{k^\prime}|,z) P_{\delta\delta}^{\rm{lin}}(k^\prime,z) \\
 \label{OV}
& &\qquad\qquad\qquad\times
\frac{k(k - 2k^\prime\mu^\prime)(1-{\mu^\prime}^2)}{{k^\prime}^2(k^2 +
k^\prime-2kk^\prime\mu^\prime)}.
\end{eqnarray}
The OV spectrum provides a useful check of the numerical simulation and
the way we correct for the missing velocity. In
the left panel of Figure \ref{fig:missing power corrected Pq}, we show
an excellent 
agreement between the OV spectrum and the simulation result at $z=9$, after
correcting for the missing velocity power due to a finite box size of
the simulation.

Finally, one can incorporate the effect of inhomogeneous
reionization into the equation by replacing $\delta$ in Equation~(\ref{MF}) by $\chi(1+\delta)$: 
\begin{eqnarray}
\nonumber
& &P_{q_{\perp}} (k,z)\\
\nonumber
 &=& \int \frac{d^3k^\prime}{(2\pi)^3}
(1-{\mu^\prime}^2)\left[P_{\chi(1+\delta),\chi(1+\delta)}(|\bold{k}-\bold{k^\prime}|)P_{vv}(k^\prime)\right.\\
\label{Reion Pq}
& &\qquad\left.-\frac{k^\prime}{|\bold{k}-\bold{k^\prime}|}P_{\chi(1+\delta), v}(|\bold{k}-\bold{k^\prime}|)P_{\chi(1+\delta), v}(k^\prime)\right].
\end{eqnarray}
Note that we do {\it not} use this equation to compute $P_{q_{\perp}}$,
but compute $P_{q_{\perp}}$ directly from the simulation. However, we
use this equation to estimate and correct for the missing power due to a
finite box size of the simulation as described in Appendix~\ref{Sec. Correcting for the Missing Power in Simulation}. We then use the corrected
$P_{q_{\perp}}$ in Equation~(\ref{C_l}) to compute the angular power spectrum. 
As shown in the right panel of Figure \ref{fig:missing power corrected Pq},
the effect of reionization inhomogeneity substantially boosts the power spectrum
relative to the homogeneously-ionized case,
while correcting for the missing velocity power of the finite 
simulation volume boosts it even further.

\section{Reionization Simulation}

\subsection{Basic simulation parameters}

The simulations that we shall use in this paper consist of two parts: 
(1) cosmological $N$-body simulations of collisionless particles using 
the ``CubeP$^3$M'' $N$-body code \citep{2012arXiv1208.5098H}; and (2)
radiative-transfer of H-ionizing photons 
in the density and source fields created from this N-body simulation results using the ``C$^2$-Ray'' (Conservative, Causal Ray-tracing) code \citep{mellema/etal:2006a}.
The details of
the simulations that we shall use in this paper are described in
\citet{iliev/etal:2011} and \citet{ahn/etal:2012}.

Unless specified otherwise, the reionization simulations are run on the density and source fields from the same N-body results with $3072^3$ particles in a comoving box of
$114~h^{-1}~{\rm Mpc}$ on a side. 
Halos are identified down to $10^8~M_\sun$ with at least 20 particles,
using a spherical overdensity halo finder with overdensity of 178 times
the mean cosmic density. 
One of the models uses another N-body simulation with a larger box of 
$425~h^{-1}~{\rm Mpc}$, with $5488^3$ particles, resolving halos down to $10^9~M_\sun$.
The background cosmology is based on the {\sl WMAP} 5-year data combined with
constraints from baryonic acoustic oscillations and high-redshift
Type Ia supernovae \citep[$\Omega_M = 0.27, \Omega_\Lambda  = 0.73, h =
0.7, \Omega_b 
= 0.044, \sigma_8 = 0.8, n_s = 0.96$;][]{komatsu/etal:2009}. 

For the $114~h^{-1}~\rm{Mpc}$, we then calculate the IGM density field from the particle data
with halos excluded adaptively-smoothed 
on to a $256^3$ radiative-transfer grid in order to generate ionization
maps using the C$^2$-Ray code. Therefore, the final
physical length resolution of the reionization models is
$d_{\rm{cell}}=0.45~h^{-1}~{\rm{Mpc}}$. The highest $l$-mode that we can
calculate from the simulation is given by
$l_{\rm{limit}}=k_{\rm{Nyq}}s(z_{\rm{ov}})$, where
$k_{\rm{Nyq}}=\pi/(2d_{\rm{cell}})$ is the Nyquist frequency, and
$s(z_{\rm{ov}})$ is the comoving distance out to the end of
reionization. 
For example, $z_{\rm{ov}}=6.6$ gives $l_{\rm{limit}}=22000$.

The new simulations also incorporate the
effects of even smaller halos in
 $10^5~M_\sun<M<10^8~M_\sun$, using a sub-grid prescription calibrated by
smaller-box N-body simulations with higher-resolution having $1728^3$ particles
in a box of $6.3~h^{-1}~{\rm Mpc}$
\citep{ahn/etal:2012}. Specifically, we find that there is a 
correlation between the number of these small-mass halos in each cell
and the total matter density averaged over that cell, with cells of size $0.45~h^{-1}~{\rm Mpc}$, which coincides with the size of the radiative transfer cells in our
$114~h^{-1}~{\rm Mpc}$ C$^2$-ray simulations. We then
use this correlation to calculate the number of small-mass halos in each
of the radiative-transfer cells in our $114~h^{-1}~{\rm Mpc}$ simulations.

For our most recent simulation, in a box $425~h^{-1}~\rm{Mpc}$ on a side, 
the RT grid has $504^3$ cells, so $d_{\rm{cell}}=0.84~h^{-1}~\rm{Mpc}$, 
slightly larger than that for the other simulations, and $l_{\rm{limit}}\sim 12000$.
In this larger-box simulation, low-mass halos between $10^8$ and $10^9~M_\sun$ are
included by a subgrid model like that described above for MHs.

\subsection{Varying physics of reionization}

What kind of sources are responsible for reionization?
In this section, we consider a set of reionization simulations based on source models of increasing sophistication from the one with only high-mass sources to the one with all kinds of sources down to least massive halos in our models.

For each halo identified in our simulation, we calculate the number of
ionizing photons which escape from it into the IGM per unit time, 
$\dot{N}_\gamma$, which is assumed to be proportional to the halo
mass, $M$:
\beq
\dot{N}_\gamma=\frac{f_\gamma M\Omega_b}{\Delta t\, \Omega_0 m_p}\,, 
\eeq
where $m_p$ is the proton mass, $\Delta t$ is the duration of each 
star-forming episode (i.e. which corresponds in practice to the radiative
transfer simulation time-step), and 
$f_\gamma = f_{\rm esc} f_\star N_\star$ 
is the number of ionizing photons produced and released by the halo
over the lifetime of the
stars which form inside it in this time step, per halo atom, 
if $f_*$ is the fraction of the halo atoms which form stars during
this burst,
$f_{\rm esc}$ is the fraction of the ionizing photons produced by
these stars which escapes into the IGM and the integrated number of
ionizing photons released over their lifetime per stellar atom is given
by $N_\star$.  The latter parameter depends on the assumed IMF for the
stellar population and can range from $\sim 4,000$ 
(e.g. for Pop II stars with a Salpeter IMF) to $\sim 100,000$ (e.g. for
a top-heavy IMF of Pop III stars).  Halos were assigned different
efficiencies according to their mass, grouped according to whether
their mass was above (``HMACHs'') or below (``LMACHs'') 
$10^9 M_\odot$ (but above $10^8 M_\odot$, the minimum resolved halo 
mass). Low-mass sources are assumed to be suppressed within 
ionized regions (for ionization fraction higher than 10\%), through 
Jeans-mass filtering, as discussed in \citet{iliev/etal:2007b}.

In addition to the source efficiency parameter, 
$f_\gamma$,  
we also define a slightly different factor, $g_\gamma$, that is 
given by
\beq
g_\gamma=f_\gamma\left(\frac{10 \;\mathrm{Myr}}{\Delta t}\right)\,
\label{eqn:g_gamma_convention}
\eeq
where ${\Delta t}$ is the time between two snapshots from the 
N-body simulation. The new factor $g_\gamma$ reflects the fact that
a given halo has a luminosity which depends on the ratio of $f_\gamma$
to ${\Delta t}$, so $g_\gamma$ has the advantage 
that it is independent of the length of the time interval 
between the density slices, and as such it allows a direct 
comparison between runs with different $\Delta t$. For the reader's 
convenience, we listed the values of both parameters in 
Table~\ref{RT parameters}. The specific 
numerical values of the efficiency parameters are strongly 
dependent on the background cosmology adopted and the minimum 
source halo mass. Therefore, parameter values for simulations 
based on different underlying cosmology and halo mass resolution should 
not be compared directly, but require cosmology and 
resolution-dependent conversion coefficients to achieve the 
same reionization history. 

\begin{figure*}
  \begin{center}
    \includegraphics[scale=1.0]{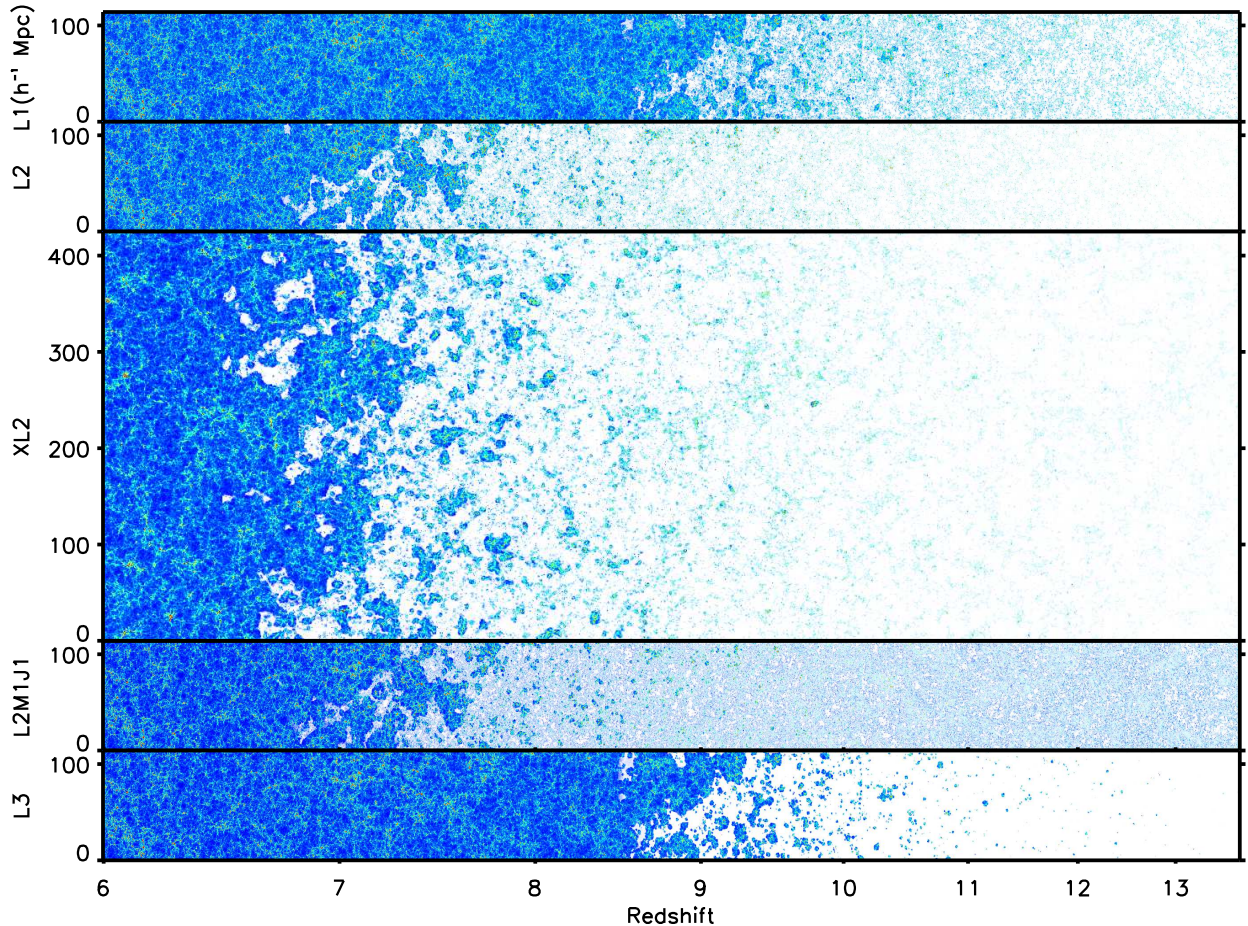}
  \caption{Cuts through the $N$-body+Radiative Transfer
   simulations used in this work. See Table~\ref{RT
   parameters} for the parameters of models L1, L2, L2M1J1, and
   L3. While these runs have the box size of $114~h^{-1}~{\rm Mpc}$, the model XL2 has
   the box size of $425~h^{-1}~{\rm Mpc}$ and has the same model parameters as
   the model L2.  Each panel shows the matter density distribution {\it
   multiplied by spatially-varying ionization fractions}. For example,
   it just shows the matter density when a given region is fully
   ionized, while it shows nothing (i.e., white) when a given region is
   fully neutral. 
   The density fields are color-coded such that overdense regions are
   red and underdense regions are blue.
   We create this figure by
   interpolating between adjacent snapshots at a given lookback time.  
 The length scale is linear in the co-moving units. 
The $x$-axis shows redshifts, while the $y$-axis shows $h^{-1}~{\rm Mpc}$.}
  \label{fig:all}
  \end{center}
\end{figure*}

\begin{figure*}
  \begin{center}
    \includegraphics[scale=0.8]{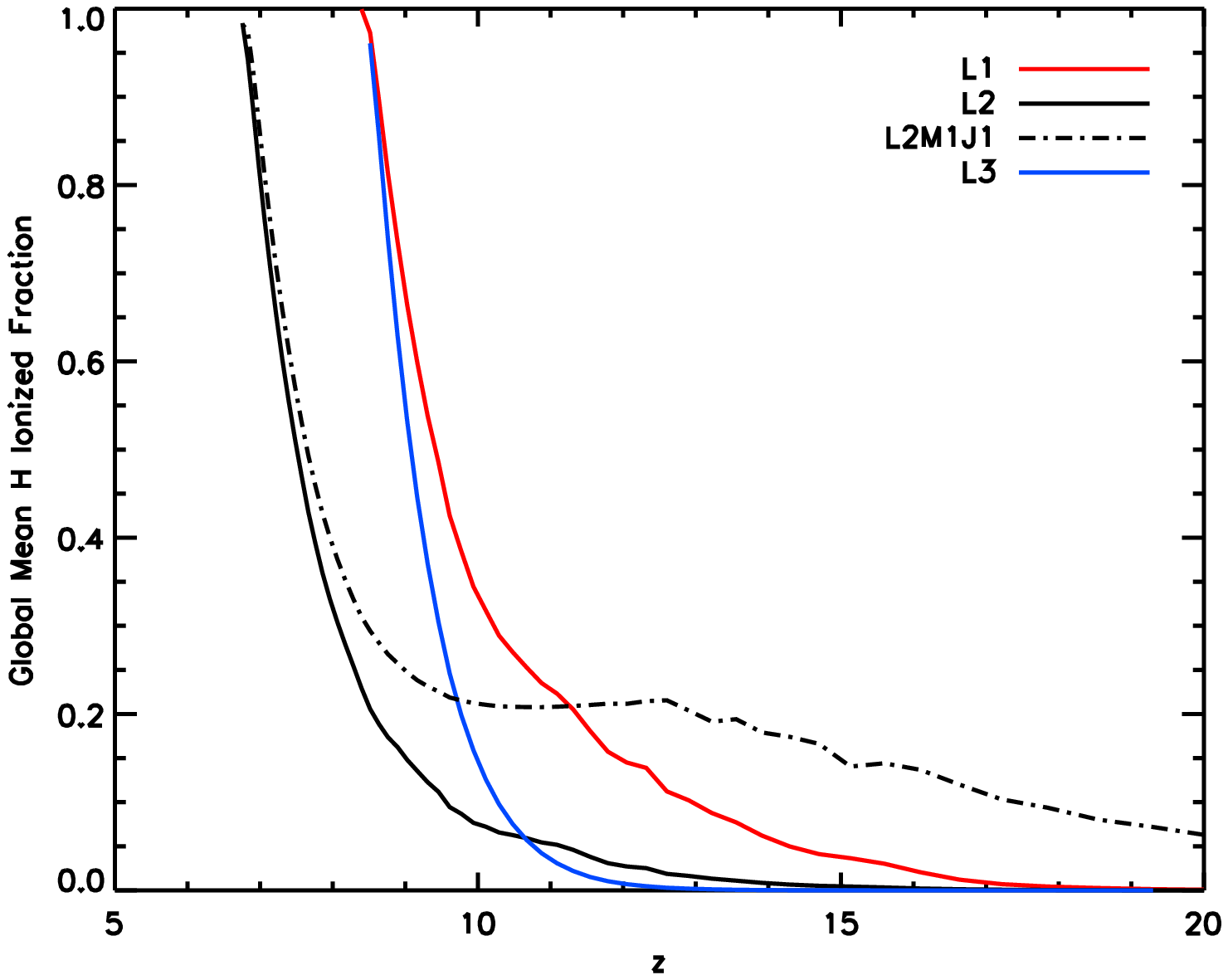}
  \caption{The global mean ionization history of our models (see Table~\ref{RT
   parameters} for the parameters of models).
   The  mass-averaged hydrogen ionization fraction, $\bar{X}$, is
   plotted against $z$. Note how self-regulation results in an
   extended period of low-level ionization by comparing the case without
   self-regulation (L3 = HMACHs only) and that with self-regulation (L1 = HMACHs + LMACHs) \citep{iliev/etal:2011}. A further extension occurs when MH sources are included, as well (i.e. compare L2 = HMACHs + LMACHs and L2M1J1 = L2 + MHs) \citep{ahn/etal:2012}.} 
 \label{fig:xmean_vs_z}
  \end{center}
\end{figure*}

\subsubsection{HMACHs-only model}

In our simplest model (labeled as L3; see Table~\ref{RT parameters}
for the details. Note that ``L'' stands for a ``large volume''), 
we only use HMACHs as the sources of reionization.
These sources are defined as the
halos with $M>2.2\times 10^9 M_{\odot}$ for L3; and with
$M>10^9~M_\sun$ for the other configurations.
These sources are believed to form stars
even when immersed in ionized regions, 
due to the fact that their gravitational potential wells are deep enough to 
overcome Jeans-mass filtering.

\subsubsection{HMACHs+LMACHs models}

What about smaller-mass halos? LMACHs are more abundant; however,
if they form inside the regions that have already been ionized, they
would not act as sources of ionizing photons. This is because
ionization heats the gas and makes its pressure too high for the gas to
collapse into such small halos \citep[and references therein]{iliev/etal:2007b}. 

When we include LAMCHs and account for this ``self-regulation'' of reionization, 
 we give LMACHs a higher efficiency, $g_\gamma$, than for HMACHs, as
 presumably it is easier for ionizing photons to escape from LMACHs than
 from HMACHs, and Pop III stars with a top-heavy initial
 mass function (IMF), which are capable of producing more
 ionizing photons than Pop II stars with a Salpeter IMF, are more likely
 to form in LMACHs. 

There are two cases which have both HMACHs and LMACHs, and we
shall call them L1 and L2. For L1, the efficiency parameter, $g_\gamma$,
is chosen such that the overlap redshift, $z_{\rm ov}=8.3$, is
similar to that of L3, $z_{\rm ov}=8.4$ (see Table~\ref{RT
parameters}). For L2, $g_\gamma$ is chosen such that $z_{\rm ov}$ is
between 6 and 7, as suggested by the quasar absorption line
observations.

For L2, we have another run with a much larger volume ($425~h^{-1}~{\rm
Mpc}$) with $504^3$ of radiative-transfer grids.
Although it does not resolve LMACHs,
we include LMACHs as a sub-grid model using correlation between 
average density of radiative transfer cells and number density of LMACHs
similarly to how \cite{ahn/etal:2012} included MHs in the simulation (Iliev et al. and Ahn et al. in preparation).
 This run gives
$l_{\rm{limit}}\sim12000$. 
We shall call this configuration
``XL2'', as the volume for this run is bigger (hence the name, XL)
than those runs with ``L.'' This run will be used to check our method to
correct for the missing velocity power.

\subsubsection{HMACHs+LMACHs+MHs model}

What about {\it even} smaller-mass sources? Gas in halos of masses between
$10^5~M_\sun$ and $10^8~M_\sun$ is thought to cool via
rotational and vibrational transitions of hydrogen molecules and form stars, until
hydrogen molecules 
are dissociated by Lyman-Werner photons in the UV background 
from other sources
\citep[see][and references therein]{ahn/etal:2012}.

The MHs form earlier than LMACHs or HMACHs, and thus can start
reionization of the universe earlier. However, as the star formation in
MHs is vulnerable to Lyman-Werner photons, it gets suppressed
wherever the intensity of the LW background rises above the threshold for suppression,
locally at first, and eventually globally. 
This adds another kind of ``self-regulation'' to the reionization history, 
with an even more extended phase of low-level ionization before MHs are eventually
suppressed completely
\citep{ahn/etal:2012}. 

The effects of MHs have been added to L2 by \cite{ahn/etal:2012}, and we
take one of the cases simulated there, L2M1J1, as our fiducial case with MHs.
See Table~\ref{RT parameters} for the efficiency of MHs. 
``M" denotes the mass spectrum of Pop III stars in MHs, and ``J" the threshold
intensity of the Lyman-Werner photon background, above which the star
formation in MHs is suppressed. In L2M1J1, each halo is assumed to host one Pop III star with mass of $300~M_\sun$, and the assumed LW threshold is $J_{\rm{LW,th}} =
0.1\times 10^{-21} ~\rm{ergs}^{-1} \rm{cm}^{-2} \rm{sr}^{-1}$. 

\section{Results} 

\begin{figure*}
  \begin{center}
    \includegraphics[scale=0.8]{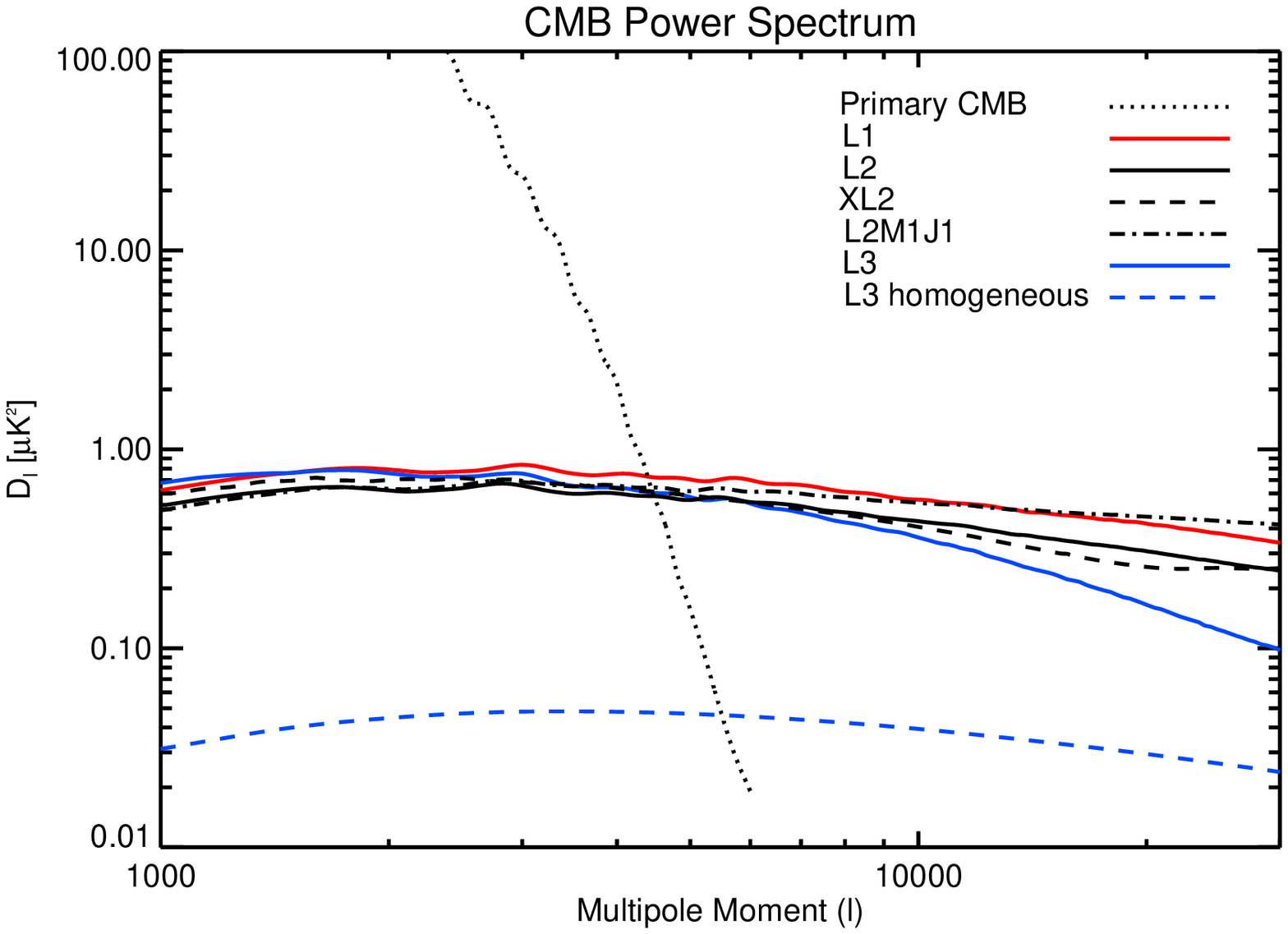}
  \caption{Predicted kSZ power spectra, $D_l^{\rm{kSZ}}$, for the
   models discussed in this work (see Table~\ref{RT
   parameters} for the parameters of models). 
   The box size of L1, L2, L2M1J1 and L3 is $114~h^{-1}~{\rm Mpc}$,
   while that of XL2 is $425~h^{-1}~{\rm Mpc}$. The model parameters of XL2 are the
   same as those of L2, and thus XL2 provides a useful check of the way
   we correct for the missing velocity power in $114~h^{-1}~{\rm Mpc}$-box
   simulations (see Appendix B for details). The primary CMB power
   spectrum is also  shown.}
 \label{fig:kSZ}
  \end{center}
\end{figure*}

Before presenting and discussing our predictions for the kSZ power
spectrum, let us 
briefly  comment on the global ionization history of the universe, which
is the key to understanding the difference between our results and the
previous ones. For more detailed discussion on the effects of
self-regulation, see 
\citet{iliev/etal:2007b,iliev/etal:2011} and \citet{ahn/etal:2012}.

Figure~\ref{fig:all} shows how the reionization proceeds in our
simulation boxes, while Figure~\ref{fig:xmean_vs_z} shows the mass-averaged
ionization fraction of the universe as a function of redshift. Both
figures show that inclusion of low-mass halos (LMACHs and MHs), which are
self-regulated, significantly extends the ionization history of the
universe toward higher redshift. 
Let us compare L1 and L3. As LMACHs form earlier, the universe
begins to be ionized earlier in L1 than in L3. However, the universe
does not get reionized quickly but keeps a low level of ionization for an
extended period due to self-regulation of sources. 
Only after HMACHs start to dominate, at $z\sim10$, 
does reionization proceed rapidly and finishes soon thereafter. 
In L3, with no LMACHs, by contrast, reionization proceeds rapidly from beginning to end because the abundance of HMACHs, the only sources, grows exponentially without 
any suppression effects to self-regulate them. 
When MHs are included (L2M1J2), the universe begins to be ionized
even earlier than the cases with HMACHs and LMACHs, and keeps a low-level ionization for a longer period. 

These physically motivated yet somewhat complex reionization histories 
were not considered in
any of the previous calculations of the kSZ power spectrum. 
In this section, we show that it is these new features 
in the reionization history that invalidate simple two-parameter
descriptions of the amplitude of the kSZ power spectrum proposed by the
previous study
\citep{zahn/etal:2012,mesinger/mcquinn/spergel:2011,battaglia/etal:2012b}.  

\subsection{Impact of Inhomogeneous Reionization}

First, it is useful to understand how important it is to
include inhomogeneity (or patchiness) of reionization when computing the
kSZ power spectrum.
In order to see this, we create a homogeneous version of L3
(``L3-homogeneous''), in which we wipe out inhomogeneity of reionization by 
replacing the ionization fraction, $\chi$, with its global average,
$\bar{\chi}$ (see Figure \ref{fig:xmean_vs_z}). 
This then gives the transverse momentum power spectrum as
$P_{q_{\perp}} = \bar{\chi}^2 P^{\rm{OV}}_{q_{\perp}}$, where
$P^{\rm{OV}}_{q_{\perp}}$ is the OV spectrum given by
Equation~(\ref{OV}). 
We remind reader that, on the scales of interest to us 
in this power spectrum $(k\lesssim 1~h~\rm{Mpc}^{-1})$,
the degree of non-linearity of the underlying density and velocity fields of the IGM is 
small enough that we can well approximate the kSZ power spectrum for this
``homogeneous'' ionization case by the assumption of linear perturbations
inherent in Equation~(\ref{OV}) 
(see Section \ref{Sec. momentum power} and 
the left panel of Figure \ref{fig:missing power corrected Pq}).
We use this momentum power spectrum in
Equation~(\ref{C_l}) to obtain the kSZ power 
spectrum for ``L3-homogeneous.'' Thus, ``L3'' and ``L3-homogeneous''
have exactly the same average reionization history, while spatial
fluctuations of ionization fraction are included only in L3. 
We find that L3 yields an order-of-magnitude larger power spectrum than L3-homogeneous 
that is consistent with findings in \cite{iliev/etal:2007}(see Figure \ref{fig:kSZ}). 

\begin{figure*}
  \begin{center}
    \includegraphics[scale=0.9]{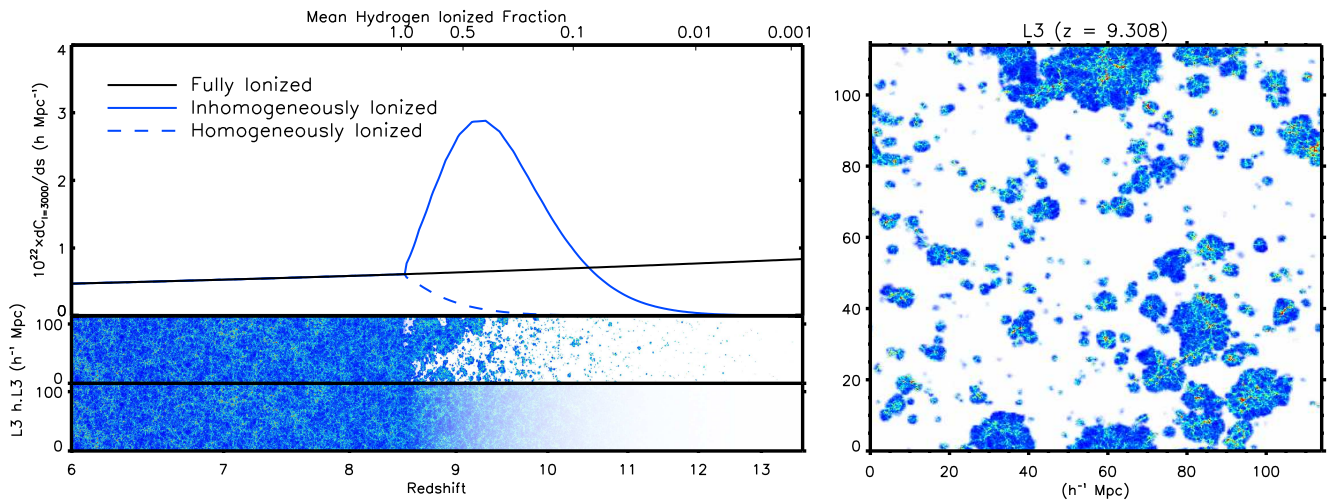}
  \caption{Left: The top panel shows the contribution from a
   given comoving distance to the kSZ 
   power spectrum at $l=3000$, $dC^{\rm{kSZ}}_{l=3000}/ds$. The solid
   line with a peak shows L3, the dashed line shows L3-homogeneous, and
   the nearly-horizontal solid line shows the fully-ionized case. The
   middle panel is the same as the bottom panel of
   Figure~\ref{fig:all}. The bottom panel shows L3-homogeneous, i.e.,
   the density distribution multiplied by the average ionization
   fraction. 
  Right: A snapshot of L3 at $z=9.3$, which gives the maximum
   contribution to the kSZ power spectrum at $l=3000$.}
 \label{fig:homo_vs_inhomo}
  \end{center}
\end{figure*}

In order to see the effect of inhomogeneous reionization on
the kSZ power spectrum in more detail, we show the contribution from a
given comoving distance to the kSZ power spectrum at $l=3000$,
$dC^{\rm{kSZ}}_{l=3000}/ds$, in Figure \ref{fig:homo_vs_inhomo}. While
both L3 and L3-homogeneous converge to the same
$dC^{\rm{kSZ}}_{l=3000}/ds$ after the universe becomes fully ionized, we find
a clear enhancement of the power when the ionization fraction is less
than unity, $z>z_{\rm ov}=8.4$. The maximum contribution occurs when the
universe is half ionized. One can see this visually in the middle (L3) and
bottom (L3-homogeneous) panels of Figure \ref{fig:homo_vs_inhomo}: L3 is
clearly more patchy than L3-homogeneous.

\begin{figure*}
  \begin{center}
    \includegraphics[scale=0.9]{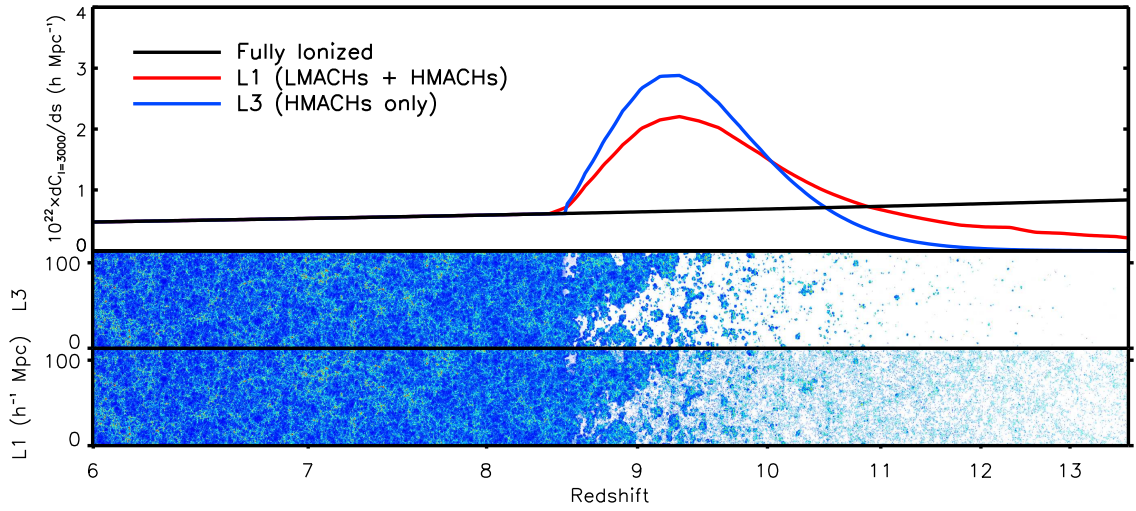}
  \caption{Same as the left panel of
   Figure~\ref{fig:homo_vs_inhomo}, but for comparing L1 (bottom panel) and L3
   (middle panel). See 
   Table~\ref{RT parameters} for the parameters of L1 and L3.}
 \label{fig:L1_vs_L3}
  \end{center}
\end{figure*}

\begin{figure*}
  \begin{center}
    \includegraphics[scale=0.6]{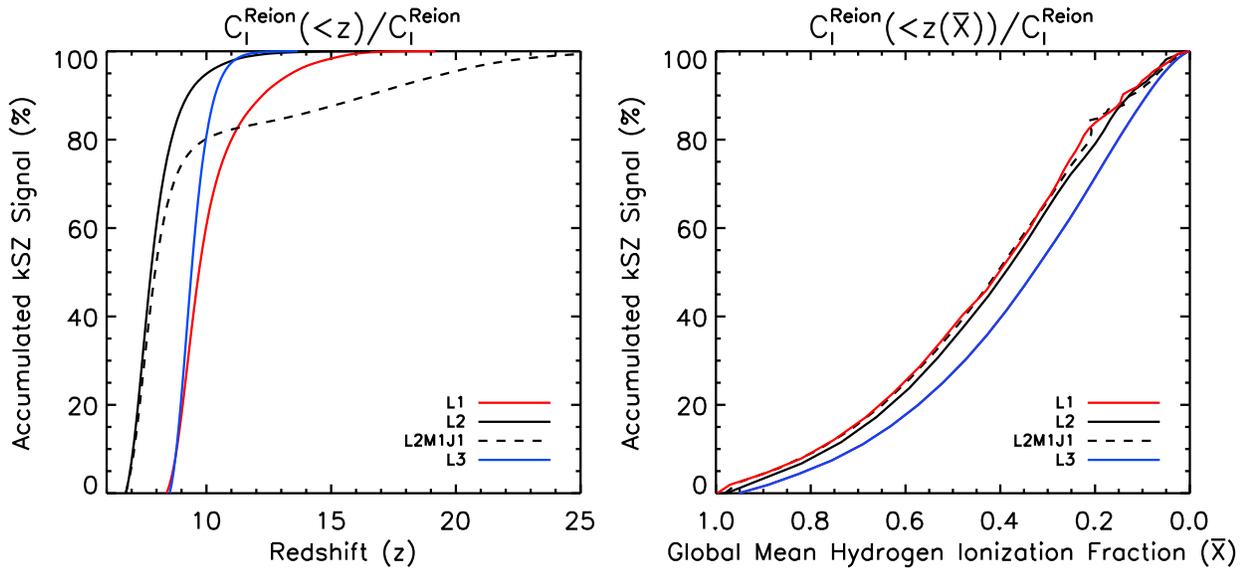}
  \caption{Cumulative reionization kSZ power spectrum at $l=3000$ as a
   function of the maximum redshift (Left) and the mean ionization
   fraction (Right).} 
 \label{fig:percentage}
  \end{center}
\end{figure*}

\begin{figure*}
  \begin{center}
    \includegraphics[scale=1]{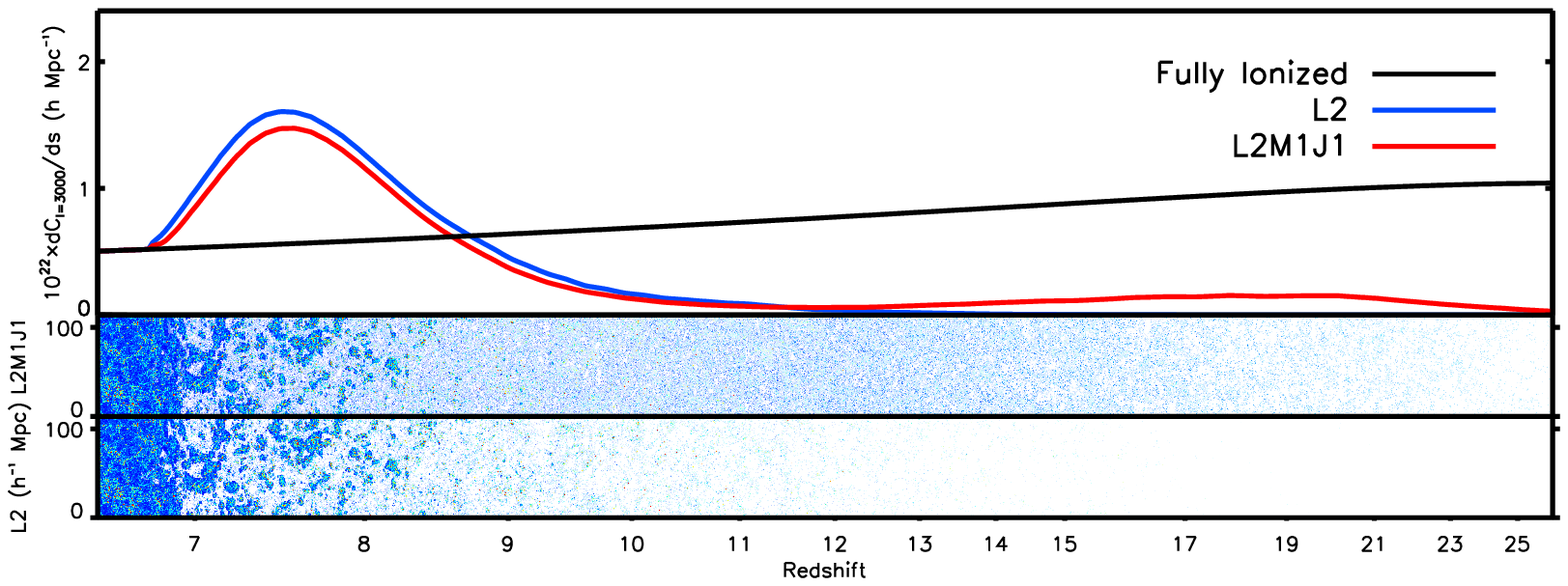}
  \caption{Same as the left panel of
   Figure~\ref{fig:homo_vs_inhomo}, but for comparing L2 (bottom panel)
   and L2M1J1 (middle panel). See  
   Table~\ref{RT parameters} for the parameters of L2 and L2M1J1.}
 \label{fig:L2_vs_L2M1J1}
  \end{center}
\end{figure*}

The angular scale for $l=3000$ roughly corresponds to the co-moving
length of $15~h^{-1}~{\rm Mpc}$ during the reionization era ($z \sim
10$). The contribution to the kSZ power spectrum continues
to grow until the typical comoving size of ionized bubbles reaches $15~h^{-1}~{\rm Mpc}$. In our models, this occurs when the universe is
half ionized. After this epoch bubbles grow bigger than $15~h^{-1}~{\rm Mpc}$, and
thus the ionization field is no longer patchy on the scale of $15~h^{-1}~{\rm Mpc}$. This
explains why the contribution to the kSZ power spectrum at $l=3000$
decreases after the half-ionization epoch. 
(By the same token, a plot like that for the inhomogeneous case L3 in 
Figure \ref{fig:homo_vs_inhomo} but for $l>3000$ 
would look similar but with the peak shifted to higher $z$, 
when ionized patches were smaller-scale.)

\subsection{Impact of LMACHs} \label{sec:LMACHs}

How does the presence of LMACHs and self-regulation affect the kSZ power
spectrum? To answer this we compare L1 and L3, which are mostly similar
except that L1 has low-mass halos ($10^8~M_\sun <M<2.2\times10^9~M_\sun$)
with most of them being LMACHs. While they finish
reionization at nearly the same redshift, L1 begins ionization earlier
due to LMACHs and gives an extended period of low ionization due
to self-regulation (see Figure~\ref{fig:xmean_vs_z}).

Figure~\ref{fig:kSZ} shows that L1 and L3 give similar kSZ power
spectra at $l\lesssim 3000$, while at higher multipoles L1 becomes
significantly greater than L3. This is because there are numerous
ionized bubbles created by LMACHs at high redshifts, which give
significant contributions to the small-scale kSZ power spectrum.
Although it would be a challenge for current surveys, future
measurements of $D^{\rm{kSZ}}_l$ with 10\% accuracy over a wide range of
multipoles can distinguish between the predictions of L1 and L3,
shedding light on the roles of LMACHs during the reionization.

We compare the contributions from a
given comoving distance to the kSZ power spectrum at $l=3000$,
$dC^{\rm{kSZ}}_{l=3000}/ds$, for L1 and L3 in Figure
\ref{fig:L1_vs_L3}. As expected, L1 has larger contributions at higher
redshifts ($z\gtrsim 10$) due to LMACHs. On the other hand, L3 has
larger contributions at lower redshifts ($z\lesssim 10$), as it is more
patchy due to the absence of smaller bubbles around LMACHs (see the
middle (L3) and bottom (L1) panels of Figure~\ref{fig:L1_vs_L3}). 
In L1, bubbles around LMACHs do not grow much because of self-regulation.

In the left panel of Figure~\ref{fig:percentage}, we show the
cumulative contributions to the kSZ power spectrum at $l=3000$ below a given
maximum redshift, $z$. This also shows that L1 receives larger
contributions from higher redshifts than L3: 20\% of the total power 
in L1 comes from $z>11$, while only a few percent of the total power in
L3 comes from $z>11$. Similarly, the right panel of Figure~\ref{fig:percentage}
shows that 20\% of the total power in L1 comes from when the ionization
fraction is less than 0.25, which is consistent with the ionization
history above $z=11$ shown in Figure~\ref{fig:xmean_vs_z}. This extended
tail has important implications for the interpretation of the kSZ power
spectrum, as we shall discuss in Section \ref{sec:interpretation}.

\subsection{Impact of Mini Halos} \label{sec:MHs}

What about MHs? We compare L2 and L2M1J1, which have the same efficiency
parameters for HMACHs and LMACHs, but only L2M1J1 considering MHs. 
While L2 and L2M1J1 finish reionization at almost the same
redshift, L2M1J1 begins ionization much earlier
due to MHs and gives a significantly more extended period of low ionization
due to self-regulation (see Figure~\ref{fig:xmean_vs_z}).

Figure~\ref{fig:kSZ} shows that L2 and L2M1J1 give similar kSZ power
spectra at $l\lesssim 5000$, while at higher multipoles L2M1J1 becomes
greater than L2. The reason is the same as that for L1 versus L3: there
are numerous ionized bubbles created by MHs at high redshifts, which
contribute  to the small-scale kSZ power spectrum.

While L2M1J1 begins reionization much earlier and thus has more
contribution from high redshifts to the kSZ power spectrum, the actual
magnitude of the high-redshift contribution is modest. This is because of
self-regulation: self-regulation prevents  bubbles around MHs
from growing, and thus we end up having numerous small bubbles filling
space nearly uniformly. This results in a lesser degree of patchiness,
hence a modest contribution to the kSZ power spectrum at $l=3000$. One
can see this 
visually in the middle (L2M1J1) and bottom (L2) panels of
Figure~\ref{fig:L2_vs_L2M1J1}. As a result, the situation is similar to
that for L1 versus L3: 20\% of the total power at $l=3000$
in L2M1J1 comes from $z>10$, while only 5\% of the total power in
L2 comes from $z>10$.

It is interesting that all the models with self-regulation
(L1, L2, and L2M1J1) lie on top of each other when the cumulative
contribution is shown as a 
function of the mean ionization fraction (see the right panel of
Figure~\ref{fig:percentage}), whereas the model that does not have
self-regulation (L3) is a clear outlier. Whether this is merely a
coincidence or a unique feature of self-regulation is unclear due to the
limited number of samples.

\section{Spot checking the previous constraints on the duration of reionization: more extended histories can give similar kSZ signals} \label{sec:interpretation}

What determines the amplitude of the kSZ power spectrum?
Recent studies using semi-numerical reionization models
\citep{zahn/etal:2012,mesinger/mcquinn/spergel:2011,battaglia/etal:2012b}
claim that the amplitude of the kSZ power spectrum at
$l=3000$ can be described by a two-parameter family: the
redshift of half-ionization, $z_{50\%}$, and the duration of
reionization defined as $\Delta z \equiv z_{\rm{99\%}}-z_{\rm{20\%}}$ 
\citep{zahn/etal:2012} or $\Delta z \equiv z_{\rm{75\%}}-z_{\rm{25\%}}$
\citep{mesinger/mcquinn/spergel:2011,battaglia/etal:2012b}. 
None of these studies included the effects of self-regulated
reionization, and thus the reionization histories explored in these
studies are roughly symmetric about the epoch of half-ionization.

\begin{table*}
\caption{Global reionization history and kSZ signal}
\begin{center}
    \begin{tabular}{|l|l|l|l|l|l|l|l|l|}
\hline
        Label         &   $z_{50\%}$  & $z_{\rm{99\%}}-z_{\rm{20\%}}$ &
     $z_{\rm{75\%}}-z_{\rm{25\%}}$ & $D_{l=3000}^{\rm{kSZ}}(z>5.5)$  &
     $D_{l=3000}^{\rm{kSZ}}(<z_{\rm{ov}})$\footnote{From the scaling
     relation of  \citet{shaw/etal:2011}.}  & $D_{l=3000}^{\rm{kSZ}}(>z_{\rm{ov}})$ & $D_{l=3000}^{\rm{kSZ,total}}$ \\  \hline
        L1             & 9.5 & 3.2  & 2.2&1.27 & 1.94  & 0.83 & 2.77      \\  
        L2             & 7.6  & 2.1  & 1.4&0.87 & 1.69  & 0.66 & 2.35      \\
        L2M1J1        & 7.7  & 6.5  & 2.1&0.90 & 1.69  & 0.69 & 2.38  \\  
        L3             & 9.1  & 1.3  & 0.9&1.20 & 1.96  & 0.75 & 2.71      \\  \hline
    \end{tabular}
\end{center}
\label{table:D_l}
\end{table*} 

Figure~2 of \citet{zahn/etal:2012} shows that the kSZ power
spectrum at $l=3000$ increases by a factor of two as the duration of
reionization increases from $\Delta z=2$ to 4. Figure 10 of
\citet{mesinger/mcquinn/spergel:2011} shows that, for a
half-ionization redshift of $z_{50\%}=9$,
the kSZ power spectrum at $l=3000$ increases by a factor of 1.4 as the
duration of reionization increases from $\Delta z=1.3$ to $2.6$. The
former gives a scaling of $D^{\rm{kSZ}}_{l=3000}\propto
(z_{\rm{99\%}}-z_{\rm{20\%}})$, whereas the latter gives $D^{\rm{kSZ}}_{l=3000}\propto
(z_{\rm{75\%}}-z_{\rm{25\%}})^{0.5}$, for a fixed half-ionization
redshift. More recently, using a new semi-numerical method based on a
correlation between the smoothed density field and the
redshift-of-reionization field found from radiation-hydro simulations of
\citet{battaglia/etal:2012a}, \citet{battaglia/etal:2012b} calculate the
kSZ power spectrum coming from $z>5.5$ and obtain the following scaling
relation: 
\begin{equation}
 D^{\rm{kSZ},z>5.5}_{l=3000}=2.02~\mu{\rm
  K}^2\left[\left(\frac{1+\bar{z}}{11}\right)-0.12\right]\left(\frac{\Delta
       z}{1.05}\right)^{0.47}, 
\label{eq:battaglia}
\end{equation}
where $\Delta z = z_{\rm{75\%}}-z_{\rm{25\%}}$ and $\bar{z}$ is the
mean value of the redshift-of-reionization field, which is approximately
equal to the half-ionization redshift, $z_{50\%}$.

Our predictions for $D^{\rm{kSZ}}_{l=3000}$ are summarized in
Table~\ref{table:D_l}. Among the models we have explored in
this paper, L3 (which contains 
only HMACHs and does not have self-regulation) closely matches the
scenarios explored in the above studies. Using $z_{50\%}=9.1$ and
$z_{\rm{75\%}}-z_{\rm{25\%}}=0.9$ we find for L3, Equation~(\ref{eq:battaglia})
gives $D^{\rm{kSZ},z>5.5}_{l=3000}=1.5~\mu{\rm K}^2$. This is in a reasonable
agreement with our result,\footnote{In order to
compute $D^{\rm{kSZ},z>5.5}_{l=3000}$, we calculate the contribution
from $z$ between $5.5$ and $z_{\rm ov}$ using the fully-ionized formula,
$P_{q_\perp}=P^{\rm{OV}}_{q_\perp}$, and add it to
$D^{\rm{kSZ}}_{l=3000}(>z_{\rm ov})$ shown in the seventh column of
Table~\ref{table:D_l}.} $D^{\rm{kSZ},z>5.5}_{l=3000}=1.2~\mu{\rm K}^2$.

However, the above formula significantly overestimates the
amplitude of the kSZ power spectrum for L1: Equation~(\ref{eq:battaglia}) gives
$D^{\rm{kSZ},z>5.5}_{l=3000}=2.4~\mu{\rm K}^2$, whereas we find
$D^{\rm{kSZ},z>5.5}_{l=3000}=1.3~\mu{\rm K}^2$. In other words, despite the fact
that L1 has a significantly more extended duration of reionization than
L3 (by a factor of more than two), $z_{\rm{75\%}}-z_{\rm{25\%}}=2.2$, the
amplitude of the kSZ power 
spectrum increases only by 8\%. Similarly, Equation~(\ref{eq:battaglia})
gives  $D^{\rm{kSZ},z>5.5}_{l=3000}=1.5$ and $1.9~\mu{\rm K}^2$ for L2
and L2M1J1, respectively, whereas we find $0.9~\mu{\rm K}^2$ for both
cases. Therefore, we conclude that Equation~(\ref{eq:battaglia}) is
valid only for simple scenarios where the reionization history is
roughly symmetric about the half-ionization redshift, but is invalid
when self-regulation is included. Similar conclusions apply to
\citet{zahn/etal:2012} and \citet{mesinger/mcquinn/spergel:2011}.

Our results show that self-regulation makes the duration of
reionization significantly more extended without changing the amplitude
of the kSZ power spectrum very much. In other words, an extended period
of low-level ionization in $z>z_{50\%}$ does not make much contribution
to the kSZ power spectrum at $l=3000$.

\section{Conclusion} 

In this paper, using the state-of-the-art reionization simulations
incorporating the effects of self-regulated reionization
\citep{iliev/etal:2011,ahn/etal:2012}, we have 
computed the power spectrum of the kSZ effect from the EOR. 
Unlike the previous work
which created maps and computed two-dimensional Fourier transforms from
the maps, we have computed the kSZ power spectrum from a line-of-sight
integral of the transverse momentum power spectrum of ionized gas.
We present a method to statistically correct for the missing
velocity power in Appendix~\ref{Sec. Correcting for the Missing Power in
Simulation}, and verify the accuracy of our method by comparing the
results from large- (425~${\rm{Mpc}}/h$) and small-box
(114~${\rm{Mpc}}/h$) simulations.

We find that the kSZ power spectrum is a sensitive probe of patchiness
of reionization: patchiness increases the amplitude of the
kSZ power spectrum by an order of magnitude. The maximum contribution
occurs when the 
angular sizes of ionized bubbles are close to those corresponding to a
given multipole.

While inclusion of small-mass halos such as LMACHs and MHs
makes the beginning of reionization earlier, self-regulation
significantly slows down the progress of reionization
\citep{iliev/etal:2007b,iliev/etal:2011,ahn/etal:2012}. This results in an extended
period of low-level ionization before more massive HMACHs dominate and
finish reionization. We find that such an extended period of low-level
ionization does not, however, make much of a contribution to the kSZ power spectrum at
$l=3000$: $D^{\rm{kSZ}}_{l=3000}$ changes only by $\sim 10\%$ despite the fact
that the duration of reionization increases by a factor of more than
two.

Our results qualitatively change the conclusions reached by
the previous work which did not include self-regulation. Recent work
\citep{zahn/etal:2012,mesinger/mcquinn/spergel:2011,battaglia/etal:2012b}
assumes that  $D^{\rm{kSZ}}_{l=3000}$ can be 
adequately parameterized by the redshift of
half-ionization, $z_{50\%}$, and the duration of reionization, $\Delta
z$. While our result for the 
simplest model of reionization without self-regulation (L3) agrees with
the scaling formula of \citet{battaglia/etal:2012b}
(Equation~\ref{eq:battaglia}), our results for the models with
self-regulation do not agree with it: specifically, the amplitude of the
kSZ effect is no longer correlated well with the duration of the
reionization. This is because self-regulation gives an extended period
of low-level reionization only for $z>z_{20\%}$, while the simple models
adopted by these other treatments have
a roughly symmetric reionization history about $z=z_{50\%}$, for which
a longer duration thus implies a longer period of patchy state with a
significant ionization across
$z=z_{50\%}$. Therefore, a more accurate scaling formula is required to take into
account the asymmetric reionization history typical of self-regulated
reionization.

Going beyond $l=3000$, we find that LMACHs and MHs do have a
considerable impact on the kSZ power spectrum on smaller angular
scales. For example, $D^{\rm{kSZ}}_{l=10000}$ is boosted by 60\% and
25\% when LMACHs and MHs are included, respectively. Even though
measurements of the kSZ power spectrum at $l>3000$ would be
a challenge for the moment due to contamination by extragalactic point
sources and tSZ, future multi-wavelength observations
may allow us to determine the kSZ power spectrum from the EOR
over a wide range of multipoles. 
Such measurements will provide us with valuable additional
information on the nature of the ionizing sources and the history of reionization.

How do our calculations compare with these current observational
constraints? In order to obtain the total kSZ signal from both
reionization and post-reionization contributions, we take the ``CSF''
(cooling and star formation) post-reionization
model of \cite{shaw/etal:2011} that approximately 
incorporates the Jeans-filtering of $P_{q_\perp}$ due to shock heating
in halos and in the IGM. The post-reionization kSZ signal computed from
their scaling relation and the total kSZ signal (i.e., the sum our reionization
calculation and their post-reionization calculation) are shown in the sixth
and seventh columns of Table \ref{table:D_l}, respectively.
We find that all of our predictions are consistent with the 95\%~CL
upper bound on the total signal from SPT, $D^{\rm{kSZ},\rm total}_{l=3000}<2.8~\mu
K^2$ \citep{reichardt/etal:2011}. Therefore, we conclude
that the current data are consistent with our understanding of the
physics of reionization.

\section{Acknowledgment} 
KA was supported in part by NRF grant funded by the Korean government
MEST (No. 2012R1A1A1014646,
2012M4A2026720). ITI was supported by The Southeast 
Physics Network (SEPNet) and the Science and Technology Facilities
Council grants ST/F002858/1 and ST/I000976/1. This study was supported
in part by the Swedish Research Council grant 2009-4088, U.S. NSF grants
AST-0708176 and AST-1009799, NASA grants NNX07AH09G, NNG04G177G and
NNX11AE09G, and Chandra grant SAO TM8-9009X. The authors acknowledge
the TeraGrid and the Texas Advanced Computing Center (TACC) at The
University of Texas at Austin (URL: http://www.tacc.utexas.edu), and
the Swedish National Infrastructure for Computing (SNIC) resources
at HPC2N (Ume\aa, Sweden) for providing HPC and visualization resources
that have contributed to the results reported within this paper. 

\appendix

\section{Derivation of  the power spectrum of the kSZ effect}
\label{app:cl}

\subsection{Suppression of longitudinal modes}

An important observation of the nature of kSZ is that it is
given by the {\it transverse} (vector-mode or spin-1) momentum field,
and the longitudinal contribution is suppressed.
To show this, we Fourier transform Equation~(\ref{Eq:kSZ2}):
\beq
\frac{\Delta T}{T}(\hat{\gamma}) = -\frac{\sigma_T n_{e,0}}{c}\int \frac{ds}{a(s)^2} e^{-\tau} 
\int\frac{d^3k}{(2\pi)^3} 
[\hat{\gamma}\cdot\tilde{\bold{q}}(\bold{k},s)]
e^{-i\bold{k}\cdot(s\hat{\gamma})}.
\eeq
Decomposing the momentum vector in Fourier space,
$\tilde{\bold{q}}$, into the longitudinal component, $\tilde{q}_{\parallel} \equiv
\tilde{\bold{q}}\cdot\hat{k}$, and the transverse component,
$\tilde{q}_{\perp} \equiv |\tilde{\bold{q}} -
\hat{k}(\tilde{\bold{q}}\cdot\hat{k})|$, we obtain
\beq
\frac{\Delta T}{T}(\hat{\gamma}) = -\frac{\sigma_T n_{e,0}}{c}\int \frac{ds}{a(s)^2} e^{-\tau} 
\int\frac{d^3k}{(2\pi)^3} 
\left[
x \tilde{q}_{\parallel}(\bold{k},s) + 
\cos (\phi_{\hat{q}} - \phi_{\hat\gamma}) (1-x^2)^{1/2} \tilde{q}_{\perp}(\bold{k},s)
\right]
e^{-iksx},
\eeq
where $x\equiv \hat{k}\cdot\hat\gamma$, and $\phi_{\hat{q}}$ and
$\phi_{\hat\gamma}$ are the angles that $\bold{k}$
makes with ${\tilde{\bold{q}}}$ and $\hat{\gamma}$, respectively.

If the factor $e^{iksx}$ oscillates much more rapidly than the other
quantities, the integral over $s$ will be small due to
cancellation. Recalling that $a(s)$, $\tau(s)$, and $\tilde{\bold{q}}$
all vary over the Hubble length scale, $kx$ should be much smaller than
$H/c$ in order to avoid the cancellation. Namely, either the wavelength
should be longer than the Hubble length, i.e., $k\lesssim H/c$, or the
mode should be nearly perpendicular to the line-of-sight direction,
i.e., $x\approx 0$. The former does not contribute much because the
amplitude of such a long-wavelength mode is small. Thus, only the modes
that are perpendicular to the line-of-sight direction, $x\approx0$, have
a chance to contribute to the kSZ signal. 

However, in this configuration, the longitudinal component of the
momentum field is also perpendicular to the line-of-sight, and vanishes
when taken a dot-product with the line-of-sight, i.e., $x
\tilde{q}_{\parallel}\approx 0$. Therefore, only the transverse mode
survives in the integral, giving
\beq \label{Eq:kSZ3}
\frac{\Delta T}{T}(\hat{\gamma}) = -\frac{\sigma_T n_{e,0}}{c}\int \frac{ds}{a(s)^2} e^{-\tau} 
\int\frac{d^3k}{(2\pi)^3} 
\cos (\phi_{\hat{q}} - \phi_{\hat\gamma}) (1-x^2)^{1/2} \tilde{q}_{\perp}(\bold{k},s)
e^{-iksx}.
\eeq

\subsection{Angular Power Spectrum}

Here, we follow steps similar to those in Chapter 7.3 of \citet{2008cosm.book.....W} to derive the angular power spectrum of CMB fluctuations induced by the kSZ effect.

Spherical harmonic decomposition of Equation~(\ref{Eq:kSZ3}) gives
\bea
a_{lm} &=& \int d^2\hat{\gamma} ~Y^m_l{}^*(\hat{\gamma}) \frac{\Delta T}{T}(\hat{\gamma})
\nonumber\\\nonumber\\  & = & 
-\frac{\sigma_T n_{e,0}}{c}
\int d^2\hat{\gamma} ~ Y_l^{m*}(\hat{\gamma})
\int\frac{ds}{a(s)^2}e^{-\tau}\int\frac{d^3k}{(2\pi)^3} 
\cos (\phi_{\hat{q}} - \phi_{\hat\gamma}) (1-x^2)^{1/2} \tilde{q}_{\perp}(\bold{k},s) e^{-iksx}
\nonumber\\\nonumber\\  & \equiv & 
-\frac{\sigma_T n_{e,0}}{c}
\int\frac{d^3k}{(2\pi)^3} ~
f_{lm}(\bold{k}),
\eea
where
\bea
\nonumber\\\nonumber\\  f_{lm}(\bold{k}) & \equiv & 
\int d^2\hat{\gamma} ~ Y_{l}^{m*} (\hat{\gamma})
\int\frac{ds}{a(s)^2}e^{-\tau} 
\cos (\phi_{\hat{q}} - \phi_{\hat\gamma}) (1-x^2)^{1/2} \tilde{q}_{\perp}(\bold{k},s) e^{-iksx}
\nonumber\\\nonumber\\  & = & 
\int d^2\hat{\gamma} ~ Y_l^{m*} (\hat{\gamma})
\int\frac{ds}{a(s)^2}e^{-\tau} 
\cos (\phi_{\hat{q}} - \phi_{\hat\gamma}) (1-x^2)^{1/2} \tilde{q}_{\perp}(\bold{k},s) 
\nonumber\\\nonumber\\ & &
\times 4\pi \sum_{LM} (-i)^L j_L (ks)Y_L^M (\hat\gamma) Y_{L}^{M*} (\hat{{k}}).
\eea

We first choose a convenient coordinate system in which the
$z$-direction lies on that of the mode vector, i.e., $\hat{k} =
\hat{z}$, and the azimuthal direction is the same as the direction of the momentum
vector, i.e., $\phi_{\hat{q}} = 0$. In this case,
$Y^{M*}_{L}(\hat{k})$ simplifies to $Y^{M*}_{L}(\hat{z}) = \delta_{M0}
\sqrt{\frac{2L+1}{4\pi}}$, giving 
\bea
f_{lm}(k\hat{z}) &=&
\sqrt{4\pi} \int\frac{ds}{a(s)^2}e^{-\tau} \tilde{q}_{\perp}(\bold{k},s) 
\sum_{L} (-i)^L  \sqrt{2L+1} j_L (ks) 
\int d^2\hat{\gamma} ~
Y_L^0 (\hat\gamma) ~
\cos \phi ~\sin\theta ~ 
Y_l^{m*} (\hat{\gamma})
\nonumber\\\nonumber\\  & = & 
\sqrt{\frac{8\pi^2}{3}} \int\frac{ds}{a(s)^2}e^{-\tau} \tilde{q}_{\perp}(\bold{k},s) 
\sum_{L} (-i)^L  \sqrt{2L+1} j_L (ks) 
\int d^2\hat{\gamma} ~
Y_L^0 (\hat\gamma) ~
\left[Y_1^{-1} (\hat\gamma) - Y_1^1 (\hat\gamma)\right] 
Y_l^{m*} (\hat{\gamma}),
\eea
where $\theta$ and $\phi = \phi_{\hat{\gamma}}$ determine the
line-of-sight vector as
$\hat{\gamma}=(\cos\theta\sin\phi,\sin\theta\sin\phi,\cos\phi)$.

The integral over $\hat{\gamma}$ can be computed using,
\bea
&&\int d^2\hat{\gamma} ~ 
Y^M_L({\hat\gamma}) ~
Y^\mu_\Lambda({\hat\gamma}) ~
Y^{m*}_l({\hat\gamma}) = \sqrt{\frac{(2\Lambda + 1)(2l + 1)}{4\pi(2L+1)}}
C_{l\Lambda}(L,M;m,-\mu)C_{l\Lambda}(L,0;0,0) \delta_{M,m+\mu},
\eea
where $C_{l\Lambda}(L,M;m,\mu)$ is the Clebsch-Gordan coefficient for
adding the angular momentum quantum numbers $(l,m)$ and $(\Lambda,\mu)$
and for forming  $(L,M)$. In our case, we have 
\bea
f_{l,m=\pm1}(k\hat{z}) &=&
\sqrt{2\pi(2l+1)} \int\frac{ds}{a(s)^2}e^{-\tau} \tilde{q}_{\perp}(\bold{k},s) 
\sum_{L} (-i)^L j_L (ks) 
\nonumber\\\nonumber\\ & &\times
\left[
C_{l1}(L,0;\mp 1,\pm 1) C_{l1}(L,0;0,0) - 
C_{l1}(L,0;\pm 1,\mp 1) C_{l1}(L,0;0,0)
\right].
\eea
Thus, the relevant coefficients are
\bea
& &C_{l1}(l+1,0;0,0) = \sqrt{\frac{l+1}{2l+1}},\qquad
C_{l1}(l+1,0;\pm1,\mp1) = \pm \sqrt{\frac{l}{2(2l+1)}},
\nonumber\\\nonumber\\ 
& &C_{l1}(l,0;0,0) = 0,\qquad
C_{l1}(l-1,0;0,0) = \sqrt{\frac{l+1}{2l+1}},\qquad
C_{l1}(l-1,0;\pm1,\mp1) = \pm \sqrt{\frac{l+1}{2(2l+1)}}.
\eea
Putting these together gives
\bea
f_{l,m=\pm1}(k\hat{z}) &=&
(-i)^{l+1} \sqrt{\frac{\pi l(l+1)}{2l+1}} \int\frac{ds}{a(s)^2}e^{-\tau} \tilde{q}_{\perp}(\bold{k},s) \left[j_{l+1}(ks) + j_{l-1}(ks)\right]
\nonumber\\\nonumber\\ &=&
(-i)^{l+1} \sqrt{\pi l(l+1)(2l+1)} \int\frac{ds}{a(s)^2}e^{-\tau} \tilde{q}_{\perp}(\bold{k},s) \frac{j_{l}(ks)}{ks}.
\eea
Now, we get back to the observer's frame by applying the standard rotation operator, $S(\hat{q})$, that takes the $z$-direction into $\hat{k}$. This gives
\beq
f_{lm}(\bold{k}) = \sum_{m^\prime=\pm 1} D^l_{m,m^\prime}(S(\hat{k})) f_{lm^\prime}(k\hat{z}),
\eeq
where $D^l_{mm'}=\langle l,m'|S|l,m\rangle$ is the matrix representation of the
finite rotation of an initial state $(l,m)$ into a final state
$(l,m')$. We obtain
\bea \label{Eq.final1}
a_{lm} &=& -\frac{\sigma_T n_{e,0}}{c}
\int\frac{d^3k}{(2\pi)^3} ~
\sum_{m^\prime = \pm 1} D^l_{m,m^\prime}(S(\hat{k}))
(-i)^{l+1} \sqrt{\pi l(l+1)(2l+1)} \int\frac{ds}{a(s)^2}e^{-\tau} \tilde{q}_{\perp}(\bold{k},s) \frac{j_{l}(ks)}{ks}.
\eea

Finally, we calculate the angular power spectrum from
$\langle a_{lm}a_{l'm'}^*\rangle=C_l\delta_{ll'}\delta_{mm'}$ and obtain
\beq
C_l = \frac{l(l+1)}{\pi} \left(\frac{\sigma_T n_{e,0}}{c}\right)^2 \int \frac{ds}{a(s)^2}e^{-\tau(s)} \int \frac{ds^\prime}{a(s^\prime)^2}e^{-\tau(s^\prime)} \int k^2 dk \frac{j_l(ks)}{ks} \frac{j_l(ks^\prime)}{ks^\prime} P_{q_{\perp}} (k,s),
\eeq 
where $P_{q_\perp}$ is the power spectrum of $\tilde\bold{q}_\perp$
defined by $(2\pi)^3 P_{q_\perp}(k)\delta^D (\bold{k} - \bold{k}^\prime)
=\langle \tilde\bold{q}_\perp
(\bold{k})\tilde\bold{q}^*_\perp(\bold{k}^\prime)\rangle$. Here, we have
used the identity,  
\beq \label{Eq.final2}
\int d^2 \hat{k} ~D^l_{m,\pm 1}(S(\hat{k})) D^{l^\prime*}_{m^\prime,\pm 1}(S(\hat{k})) = \frac{4\pi}{2l+1}\delta_{mm^\prime} \delta_{ll^\prime}.
\eeq

The integral over $k$ can be performed with Limber's approximation: when
a function $g(k,s)$ varies much more slowly than the spherical Bessel
function, one can approximate the integral as
\begin{equation}
\int k^2 dk j_l(ks) j_l(ks^\prime) g(k,s) \approx 
\frac{\pi}{2}\frac{\delta^D (s-s^\prime)}{s^2} g\left(k=\frac{l}{s},s\right).
\end{equation}
With this approximation, we finally obtain the desired formula for the
kSZ power spectrum:
\beq
C_l = \left(\frac{\sigma_T n_{e,0}}{c}\right)^2 \int \frac{ds}{s^2a(s)^4}e^{-2\tau(s)} \frac{P_{q_{\perp}} (k=l/s,s)}{2}.
\eeq
This is Equation~(\ref{C_l}).

\section{Correcting for the Missing Power in Simulations} \label{Sec. Correcting for the Missing Power in Simulation}

The transverse momentum power spectrum at a given wavenumber,
$P_{q_{\perp}}(k)$, receives contributions from the density and velocity
auto/cross power spectra at various wavenumbers via Equation~(\ref{Reion
Pq}). As a result, $P_{q_{\perp}}$ computed from a simulation with a
finite box suffers from a loss of power due to the lack of modes whose
wavelength is greater than the size of the box \citep{iliev/etal:2007}.

The missing power arises because we do not have $P_{\chi(1+\delta),
\chi(1+\delta)}(k)$, $P_{vv}(k)$, or $P_{\chi(1+\delta), v}(k)$ for
$k<k_{\rm{box}} \equiv 2\pi/l_{\rm{box}}$, where $l_{\rm{box}}$ is the
size of the box. In Equation~(\ref{Reion Pq}), this  leads to the
missing contributions in $|\bold{k}^\prime|<k_{\rm{box}}$ and
$|\bold{k}-\bold{k}^\prime|<k_{\rm{box}}$. Estimating and correcting for
the missing power thus requires the knowledge of the large-scale limit
of $P_{\chi(1+\delta),\chi(1+\delta)}$, $P_{vv}$, and $P_{\chi(1+\delta), v}$. 

For the homogeneous reionization case, it is
straightforward to recover the missing power, as the large-scale limits
of $P_{vv}$,
$P_{\chi(1+\delta),\chi(1+\delta)}(=\bar\chi^2P_{\delta\delta})$, and
$P_{\chi(1+\delta),v}(=\bar\chi P_{\delta v})$ are precisely known by the
cosmological linear perturbation theory. Using $P_{\delta\delta}$ from
the linear theory and the linear relation, $P_{vv} =
(\dot{a}f/k)^2P_{\delta\delta}$, we find that the
missing-power-corrected momentum power spectrum from the $N$-body simulation
agrees precisely with 
the expected OV spectrum (see Figure \ref{fig:missing power corrected
Pq}). Note that most of the missing power comes from
$P_{\delta\delta}(|\bold{k}-\bold{k^\prime}|)P_{vv}(k^\prime)$ in
$k^\prime<k_{\rm{box}}$ because of the relation, $\bold{v}(\bold{k})
\propto \delta(\bold{k})/k$, in the large-scale limit.

For the inhomogeneous reionization case, we do not have a precise way to
calculate the ionized density power,
$P_{\chi(1+\delta),\chi(1+\delta)}$, in the large-scale limit; however,
we expect that the density field and the ionization
field are reasonably flat at the scales larger than the box size, and
correct for the missing bulk velocity of the box.
Therefore, we expect that the term
$P_{\chi(1+\delta),\chi(1+\delta)}(|\bold{k}-\bold{k^\prime}|)P_{vv}(k^\prime)$
in $k^\prime<k_{\rm{box}}$ captures most of the missing power, as we
have seen from the homogeneous reionization case above. 
With this approximation, the missing power in the inhomogeneously ionized regime is given by
\beq \label{Missing Pxq}
P_{q_{\perp}}^{\rm{Missing}} (k,z) = \int_{k<k_{\rm{box}}} \frac{d^3k^\prime}{(2\pi)^3} 
(1-{\mu^\prime}^2)P_{\chi(1+\delta),\chi(1+\delta)}(|\bold{k}-\bold{k^\prime}|)P_{vv}(k^\prime).
\eeq
In order to check the accuracy of Equation~(\ref{Missing Pxq}), we
compare the missing-power-corrected momentum power spectrum from the box
of $114~h^{-1}~{\rm Mpc}$ (black solid line; denoted as L2) with that from a larger box of
$425~h^{-1}~{\rm Mpc}$ (black dashed line; XL2) in Figure~\ref{fig:kSZ}. We find a
very good agreement between the two, confirming the robustness of our
correction for the missing power.

\bibliographystyle{apj}

\end{document}